\documentclass[prb,english,twocolumn,amsmath,amssymb,superscriptaddress,nofootinbib]{revtex4-2}
\usepackage{epsfig, graphicx,graphics,amsmath,amssymb,float}
\usepackage[bookmarks=true,colorlinks,linkcolor=blue,urlcolor=blue,citecolor=blue]{hyperref}
\usepackage[T1]{fontenc}
\usepackage[latin9]{inputenc}
\usepackage{lineno}
\usepackage{amsmath}
\usepackage{amssymb}
\usepackage{appendix}
\usepackage{amscd}
\usepackage{bm}
\usepackage{psfrag} 
\usepackage{bbm} 
\usepackage{babel}
\usepackage{wasysym }
\usepackage{mathrsfs}
\usepackage{color}
\usepackage{xcolor}
\usepackage{bbold}
\usepackage{soul}
\usepackage{verbatim}
\usepackage{hyperref}
\usepackage{array}
\usepackage{braket} 
\newcommand{\mc}{\mathcal}
\newcommand{\mb}{\mathbf} 
\newcommand{\pd}{ {\phantom{\dagger}} }
\newcommand{\be}{\begin{equation}}
\newcommand{\ee}{\end{equation}}
\newcommand{\bea}{\begin{eqnarray}}
\newcommand{\eea}{\end{eqnarray}}
\def \tr{\mathrm{tr}}
\def \T{{\mathrm{T}}}
\def \I{{\mathrm{I}}}
\def \J{{\mathrm{J}}}
\def \im{\mathrm{i}}

\def \rmN{\mathrm{N}}
\def \rmG{\mathrm{G}}
\def \rmU{\mathrm{U}}

\def \rmV{\mathrm{V}}
\def \rmA{\mathrm{A}}
\def \rmB{\mathrm{B}}
\def \rmC{\mathrm{C}}

\newcommand{\new}[1]{\textcolor{black}{#1}}
\begin{document}

%\linenumbers
\title{Electric polarization near vortices in the extended Kitaev model}

\author{Lucas R. D. Freitas} 
\affiliation{International Institute of Physics and Departamento de F\'isica Te\'orica
e Experimental, Universidade Federal do Rio Grande do Norte, 
Natal, RN, 59078-970, Brazil}
\affiliation{Institut f\"ur Theoretische Physik,
Heinrich-Heine-Universit\"at, D-40225  D\"usseldorf, Germany}
\author{Tim Bauer}
\affiliation{Institut f\"ur Theoretische Physik,
Heinrich-Heine-Universit\"at, D-40225  D\"usseldorf, Germany}
\affiliation{International Institute of Physics and Departamento de F\'isica Te\'orica
e Experimental, Universidade Federal do Rio Grande do Norte, 
Natal, RN, 59078-970, Brazil}
\author{Reinhold Egger}
\affiliation{Institut f\"ur Theoretische Physik,
Heinrich-Heine-Universit\"at, D-40225  D\"usseldorf, Germany}
\author{Rodrigo G. Pereira}
\affiliation{International Institute of Physics and Departamento de F\'isica Te\'orica
e Experimental, Universidade Federal do Rio Grande do Norte, 
Natal, RN, 59078-970, Brazil}

\begin{abstract}
We formulate a Majorana mean-field theory for the extended $JK\Gamma$ Kitaev model in a magnetic Zeeman field of arbitrary direction, and apply it for studying spatially inhomogeneous   states   harboring vortices. 
This mean-field theory is exact in the pure Kitaev limit and captures the essential physics throughout the Kitaev spin liquid phase.
We determine the charge profile around vortices and the corresponding quadrupole tensor. The quadrupole-quadrupole 
interaction between distant vortices is shown to be either repulsive or attractive, depending on parameters.  We predict that
 electrically biased scanning probe tips enable the creation of vortices at preselected positions. 
Our results open new perspectives for the electric manipulation of Ising anyons in Kitaev spin liquids. 
\end{abstract}
\maketitle

\section*{Introduction}

A hallmark  of  Kitaev spin liquids is  the fractionalization of spin-$1/2$ local moments into Majorana fermions and a $\mathbb Z_2$ gauge field \cite{Kitaev2006,Savary2017,Zhou2017rev,Trebst2022,Winter2017rev,Takagi2019,Hermanns2018,Knolle2019,Motome2020}. 
When time reversal symmetry is broken by an external magnetic field, both types of excitations become gapped, and vortices of the $\mathbb Z_2$ gauge field bind Majorana zero modes that behave as non-Abelian anyons. These properties can be demonstrated in the exactly solvable Kitaev honeycomb model  \cite{Kitaev2006}.  Since the observation that the bond-directional exchange interactions of the pure Kitaev model are realized in quasi-two-dimensional Mott insulators with strong spin-orbit coupling \cite{Jackeli2009}, identifying signatures of fractional excitations in  Kitaev materials has become a major goal of condensed matter physics \cite{Plumb2014,Sandilands2015,Banerjee2016,Baek2017}.    Most notably, there is evidence for a half-quantized thermal Hall conductance in the candidate material $\alpha$-RuCl$_3$ at intermediate temperatures and magnetic fields, but its interpretation in terms of chiral Majorana edge modes remains controversial \cite{Kasahara2018,Yokoi_2021,Czajka_2022,Bruin2022}. This ambiguity calls for alternative experimental probes that may help distinguish a Kitaev spin liquid from a more conventional partially polarized phase with topological magnons \cite{Cookmeyer2018,Chern2021}. 

A promising route to detect and manipulate the fractional excitations of Kitaev spin liquids is to exploit their nontrivial responses to electrical probes.  Theoretical proposals in this direction include \new{electric dipole contributions to the subgap optical conductivity \cite{bolens2018,katsura2018},} scanning tunneling spectroscopy \cite{Feldmeier2020,Konig2020,Carrega2020,Udagawa2021,Bauer2023}, interferometry in electrical conductance \cite{Aasen2020,halasz2023gatecontrolled}, and electric polarization and orbital currents associated with localized excitations \cite{Pereira2020,Banerjee2023}. In fact, the  charge polarization in  Mott insulators can be captured by an effective   density operator written in terms of spin correlations in the low-energy sector \cite{Bulaevskii_2008,Khomskii_2010}. The  effective   density operator for Kitaev materials was derived in Ref.~\cite{Pereira2020}  starting from the multi-orbital Hubbard-Kanamori model in the ideal limit where the dominant exchange path   only generates the pure Kitaev interaction \cite{Jackeli2009}. The electric field effects  then work both ways. On the one hand, the inhomogeneous  spin correlations around a $\mathbb Z_2$ vortex imply that vortices produce an intrinsic electric charge distribution. On the other hand, vortices are attracted by electrostatic potentials that locally modify  exchange couplings, and this effect   can be used to trap and move anyons adiabatically \cite{Pereira2020,Jang2021}.   

In this work we generalize the theory of the electric charge response in Ref.~\cite{Pereira2020} to consider the generic spin model for Kitaev materials \cite{Rau2014,Rousochatzakis2023}. 
Our starting point is the three-orbital Hubbard-Kanamori model which takes into account sub-dominant hopping processes that, in addition to Kitaev ($K$) interactions, also generate Heisenberg ($J$) and off-diagonal  ($\Gamma$) exchange interactions. Using  perturbation theory to leading order in the hopping parameters, we derive an expression for the effective charge density operator in  the Mott insulating phase that contains all two-spin terms allowed by symmetry. Since the additional interactions spoil the integrability of the pure Kitaev model, we compute spin correlations using a  Majorana mean-field theory. This type of approximation has been   applied to map out the ground state phase diagram and to compute response functions of the extended Kitaev model  \cite{Knolle2018,Liang2018,Nasu2018,Gohlke2018,Go2019,Ralko2020,Li2022,Yilmaz2022,Merino2022,Cookmeyer2023}. Here we generalize the mean-field approach to treat position-dependent order parameters in the case where translation symmetry is broken by the presence of  vortices in the $\mathbb Z_2$ flux configuration. Including a Zeeman coupling, we show that the spatial anisotropy of the charge distribution around a vortex varies with the direction of the magnetic field and can be quantified by the components of the electric quadrupole moment. We also discuss how a local electrostatic potential renormalizes the couplings in the extended Kitaev model and gives rise to an effective attractive potential for vortices. Remarkably, the effect is stronger in the presence of non-Kitaev interactions, and we find that it is  possible to close the vortex gap by means of electric modulation of the local spin interactions.  

\section*{Results}

\subsection*{Mean-field theory for the extended Kitaev model}

The local degrees of freedom of Kitaev materials are transition metal ions with $4d^5$ or $5d^5$ electronic configuration and strong spin-orbit coupling \cite{Trebst2022,Winter2017rev}. In the presence of the crystal field of an octahedral ligand cage, this configuration is equivalent to a single hole in a $t_{2g}$ orbital.
Starting from a three-orbital Hubbard-Kanamori Hamiltonian on the honeycomb lattice, 
in the presence of a Zeeman coupling 
to an external magnetic field $\mb h$, we find from a projection scheme 
that the low-energy effective spin Hamiltonian is given by the 
extended Kitaev (aka $JK\Gamma$) model \cite{Rau2014},
 \begin{equation}
     H =\frac{1}{2}\sum_{ ij }\sum_{\alpha \beta}   \sigma_i^\alpha   \mathbf{J}^{\alpha\beta}_{ij} \sigma^\beta_j   -\sum_i \mb h\cdot \boldsymbol\sigma_i ,\label{Hwithfield}
\end{equation}
where $\boldsymbol \sigma_i$ denotes the vector of the pseudospin-1/2 Pauli operators at site $i$. Moreover,  
$i$ and $j$ are nearest neighbors, $\mathbf{J}_{ij}$ is the bond-dependent exchange matrix,  and the indices $\alpha,\beta,\gamma\in\{ x,y,z\}=\{1,2,3\}$ label both spin components and bonds on the honeycomb lattice. We denote by $\langle ij\rangle_\gamma$ a nearest-neighbor bond of type $\gamma$ with site $i$ belonging to sublattice A and $j$ to sublattice B.
For bond $\langle ij\rangle_z$, we have 
$\mathbf{J}_{\braket{ij}_z} =    \left( \begin{array}{ccc}      J& \Gamma & 0  \\
         \Gamma & J & 0 \\          0 & 0 & J +K    \end{array}\right).$
The exchange matrices for $x$ and $y$ bonds follow by cyclic permutation of the spin and bond indices. 
The ideal Kitaev case with $J=\Gamma=0$ corresponds to a single hopping path mediated by ligands on edge-sharing octahedra  with ideal $90^\circ$ bonds   \cite{Jackeli2009}. Numerical studies show that the Kitaev spin liquid phase is stable in the regime   $|\Gamma|,|J|\ll |K|$ \cite{Rau2014,Catuneanu2018,Gordon2019,Wang2019}. For estimates of the hopping  and exchange parameters for $\alpha$-RuCl$_3$, see for instance Refs.~\cite{Winter_2016,Winter2017rev}.  In this material, one finds a ferromagnetic Kitaev coupling $(K<0$) and the leading perturbation to the idealized Kitaev model is given by $0<\Gamma<|K|$. 

We employ a mean-field approximation for calculating spin correlations in the extended Kitaev model and 
to verify the stability of the spin liquid phase against integrability-breaking perturbations.
For $J=\Gamma=h=0$, the model can be solved exactly \cite{Kitaev2006} using the Kitaev representation  $\sigma_i^\gamma =  
\im  c_i^0 c_i^\gamma$ in terms of  four Majorana fermions which obey $(c_i^\mu)^\dagger=c_i^\mu$ and $\{ c^\mu_i, c^\nu_j \} = 2\delta_{ij}\delta_{\mu\nu}$. Throughout, we use indices 
$\mu,\nu,\rho\in\{0,1,2,3\}$ to denote all four fermion flavors, in contrast with  $\alpha,\beta,\gamma\in\{1,2,3\}$. 
Physical states must respect the local constraint $D_i = c^0_i c^1_i c^2_i c^3_i = +1$.  
The algebra of the spin operators can be satisfied using  different representations \cite{Fu2018}. It is convenient to write the Kitaev representation in terms of 
the vector $c_i^{\phantom 1}=(c_i^0,c_i^1,c_i^2,c_i^3)^{\rm T}$ and the antisymmetric matrices $\rmN^\gamma$ defined by 
\begin{equation}
    \sigma_i^\gamma =   \frac{\im}{2}  c_i^{\T} \rmN^\gamma  c_i \equiv \frac{\im}{2}\Big( c_i^0c_i^\gamma - c_i^\gamma c_i^0 \Big) .
    \label{eq:spin_rep}
\end{equation}
Instead of imposing $D_i=+1$, we use the equivalent constraint \cite{Seifert_2018}
\begin{align} 
    c_i^{\T} \rmG^\gamma  c_i 
    &\equiv c_i^0c_i^\gamma - c_i^\gamma c_i^0 +\sum_{\alpha\beta} \epsilon^{\alpha \beta \gamma} c_i^\alpha c_i^\beta = 0 .  \label{eq:G=0}
\end{align}
Note that the constraints $c_i^{\T} \rmG^\gamma  c_i =0$ for $\gamma=x,y,z$ are redundant. 
If the constraint is implemented exactly, it suffices to impose it for a single value of $\gamma$. However, when treating the constraints \eqref{eq:G=0} numerically through the corresponding 
Lagrange multipliers $\lambda_i^\gamma$ \cite{Ralko2020,Yilmaz2022}, it is advantageous to 
enforce them in a symmetric manner for all three values of $\gamma$.  
We thereby rewrite the spin Hamiltonian as
\bea
        H&=&\frac{1}{8}\sum_{ij } \sum_{\alpha\beta}
         \im c_i^{\T} \rmN^\alpha  c_i  \,  \mathbf{J}_{ij}^{\alpha\beta} \,  \im c_j^{\T} \rmN^\beta  c_j  \nonumber\\
         &&-  \frac{1}{4} \sum_{i\gamma} \left( 2 h^\gamma \im c_i^{\T} \rmN^\gamma  c_i - \lambda_i^\gamma \im c_i^{\T} \rmG^\gamma  c_i  \right)   .
          \label{eq:H}
\eea
We decouple the quartic terms using two types of real-valued  mean-field parameters,   
\begin{align}
         \rmU_{ij}^{\mu\nu} = \braket{\im c_i^\mu c_j^\nu } , \qquad  \rmV_{i}^{\mu\nu} &= \braket{\im c_i^\mu c_i^\nu }      , \label{eq:MF_parameters}
\end{align}
which obey   $ \rmU_{ij}^{\mu\nu} =-  \rmU_{ji}^{\nu\mu}$ and $  \rmV_{i}^{\mu\nu}  = 2 \im \delta^{\mu\nu}- \rmV_{i}^{\nu\mu}$. 
For the exactly solvable  Kitaev model, one finds   that    $\rmU^{\mu\nu}_{ij}$ is diagonal in the indices $\mu,\nu$. In particular, the components $\rmU^{\gamma\gamma}_{ij}$ are related to the static $\mathbb Z_2$ gauge field and take values $\rmU^{\gamma\gamma}_{ij}=\pm1$  when $i,j$ form a nearest-neighbor $\gamma$ bond, and  $\rmU^{\gamma\gamma}_{ij}= 0$ otherwise.   Thus, $\rmU^{\gamma\gamma}_{ij}$ can be viewed as  an   ``order parameter'' for the Kitaev spin liquid phase.   For comparison with the exact solution, we  also define 
$W_p=\prod_{\langle ij\rangle_\gamma \in p}\rmU_{ij}^{\gamma\gamma}$,
where $p$ is a hexagonal plaquette.
In the pure Kitaev model, $W_p$ is identified with the gauge-invariant $\mathbb Z_2$ flux, and the ground state lies  in the sector with  $W_p=+1$ for all plaquettes. States with  $W_p=-1$ at isolated  plaquettes are associated with  vortex excitations \cite{Kitaev2006}.  Besides the link variables $\rmU_{ij}^{\mu\nu}$, in the mean-field approach we also consider the on-site fermion bilinears $\rmV_i^{\mu\nu}$. It follows from the Kitaev representation that  $ \rmV_{i}^{0\gamma}=\braket{\sigma_i^\gamma}$.  Moreover, the constraint in
Eq.~(\ref{eq:G=0}) implies   $\rmV_{i}^{\alpha\beta}=-\rmV_{i}^{0\gamma}$ for $(\alpha\beta\gamma)$ a cyclic permutation of $(xyz)$. Thus, there are only three independent components of $\rmV_i^{\mu\nu}$ at each site, and they are related to the local magnetization induced by the external magnetic field. In the limit $|h|\gg |K|,|J|,|\Gamma|$, we expect to encounter  a partially polarized phase characterized by $ \rmV_{i}^{\mu\nu}\neq0$ while $\rmU^{\mu\nu}_{ij}= 0$ for all bonds.  For further detail, see the Methods section.

\subsection*{Homogeneous case}

\begin{figure}[t]
\begin{center}
\includegraphics[width=\columnwidth]{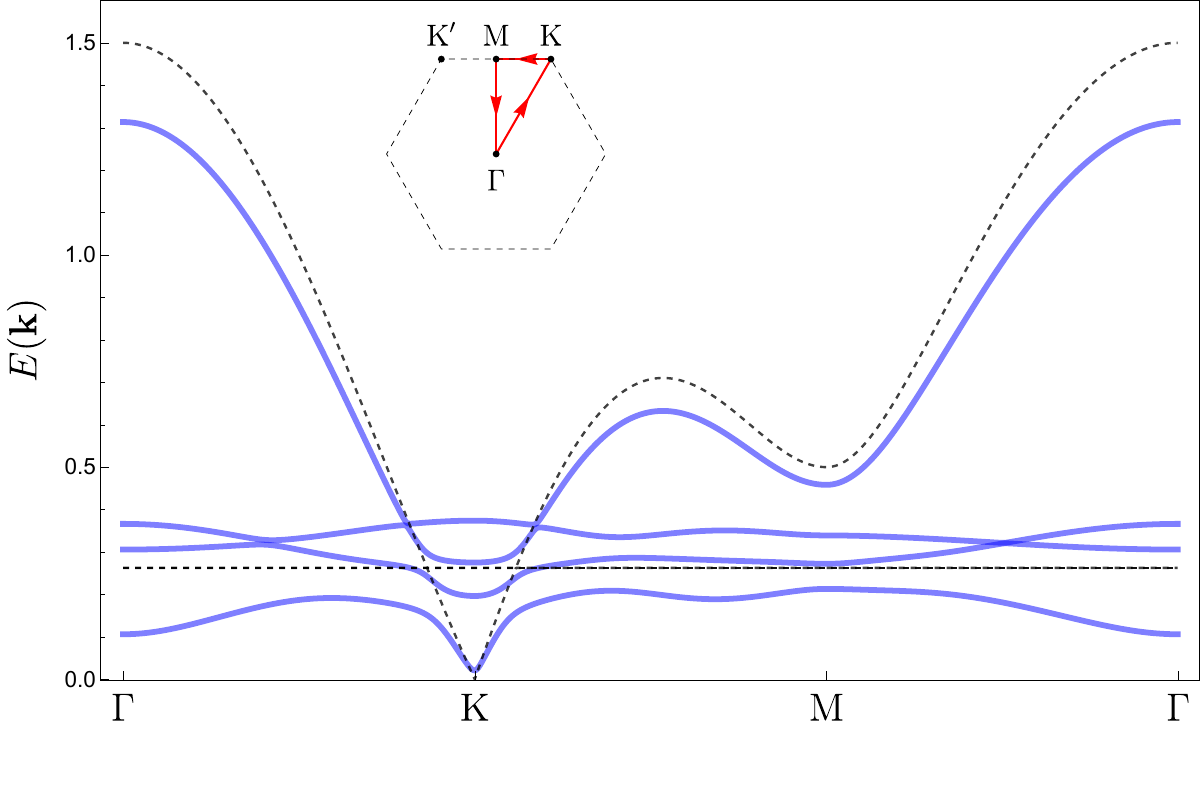}
\end{center}
\caption{Dispersion relation of Majorana fermions calculated within the mean-field approach for the homogeneous system with  $K=-1$, $J=0$, $\Gamma=  0.2$, and $\mb h=0.4\hat{\mb c}$, along the indicated BZ path.  For comparison, the dashed lines show the dispersion in the pure Kitaev limit ($\Gamma=J=h=0$). }
\label{fig1}
\end{figure}

We first describe the mean-field solution for the homogeneous case, i.e., in the absence of vortices. If the ground state does not break spin rotation or lattice symmetries, as in the Kitaev spin liquid phase, the matrices     $\rmU^{\mu\nu}_{ij}$  depend only on the bond type $\gamma$, and we set $\rmU^{\mu\nu}_{ij}=\rmU^{\mu\nu}_\gamma$ for bonds $\braket{ij}_\gamma$.  
Moreover, $\rmV^{\mu\nu}_{i}=\rmV^{\mu\nu}$ becomes a constant matrix. 
More generally, we can allow these parameters to vary with the sublattice within larger unit cells to describe magnetically ordered phases.   We then solve the mean field self-consistency equations using a Fourier transform of the Majorana modes in the thermodynamic limit. As a first step, we have verified that our mean-field approach recovers the exact results for the  
Kitaev model \cite{Kitaev2006} when we set $\Gamma=J=h=0$.    
The resulting dispersion relation of Majorana fermions is depicted by dashed lines in Fig.~\ref{fig1}. In this case, the only dispersive band is associated with the fermion $c^0$.  
This band is gapless with a Dirac spectrum near the K point in the Brillouin zone (BZ).
In addition, there are three degenerate flat bands associated with the fermions $c^\gamma$, which are related to the static gauge variables $\rmU^{\gamma\gamma}_\gamma$ (whose value is independent of $\gamma$).

\begin{figure}[t]
\begin{center}
\includegraphics[width=\columnwidth]{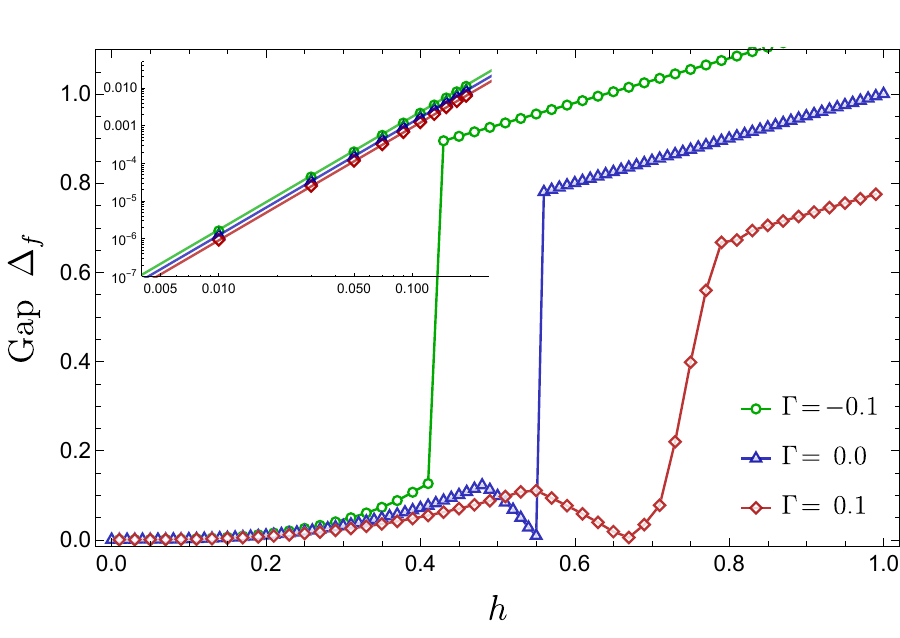}
\end{center}
\caption{Fermion gap as a function of magnetic field for $\mb h=h\hat{\mb c}$ along the $\hat{\mb c}$ axis, with $K=-1$, $J=0$, and for three values of $\Gamma$:  $\Gamma=-0.1$ (green circles), $\Gamma=0$ (blue triangles) and $\Gamma=0.1$ (red diamonds). The inset shows that for weak fields the fermion gap agrees with the perturbative result to leading order in $h$, $\Delta_f\propto h^3$. 
}
\label{fig2}
\end{figure}

Moving away from the exactly solvable point, we find that all bands become dispersive. For $h=0$ and $K, J, \Gamma\neq0$,  our results are in quantitative agreement with a previous mean-field calculation \cite{Knolle2018}. 
Our approach also allows us to take into account the magnetic field nonperturbatively.  
Figure  \ref{fig1} shows the dispersion for a magnetic field pointing along the crystallographic  $c$ direction  (perpendicular to the honeycomb plane), with unit vector $\hat{\mb c}=\frac1{\sqrt3}(1,1,1)$. Here  the coordinates are specified in terms of the crystallographic axes $\hat{\mb x}$, $\hat{\mb y}$ and $\hat{\mb z}$ of the ligand octahedra.  For later reference, the in-plane unit vectors are $\hat{\mathbf a}=\frac1{\sqrt6}(1,1,-2)$ and $\hat{\mathbf b}=\frac1{\sqrt2}(-1,1,0)$. As shown in Fig.~\ref{fig2},  the magnetic field opens up a gap in the fermion spectrum, as expected for the non-Abelian Kitaev spin liquid phase. As we increase the magnetic field, the gap at the K point increases, but the gap at the $\Gamma$ point decreases. The fermion gap  $\Delta_f$ is given by the minimum between the energies at the K and $\Gamma$ points in the BZ.  
If these energies cross, $\Delta_f$ exhibits a kink at the corresponding value of $h$ (e.g., for $\Gamma=0$ in Fig.~\ref{fig2}). As we increase the magnetic field, we encounter a critical value $h_c$ at which the gap either changes discontinuously, as in a first-order transition (e.g., for $\Gamma=-0.1|K|$ in Fig.~\ref{fig2}), or it vanishes and varies continuously across the phase transition (e.g., for $\Gamma=0.1|K|$ in Fig.~\ref{fig2}). 
 For   $\mb h=h\hat{\mb c}$ and $h\ll h_c$, the fermion gap increases with the magnetic field as $\Delta_f\propto h^3$, as expected from perturbation theory \cite{Kitaev2006}; see the inset in Fig.~\ref{fig2}. For general field directions, the fermion gap behaves as $\Delta_f\propto h_x h_y h_z$, closing when one component of $\mb h$ vanishes.

\begin{figure}[t]
\begin{center}
\includegraphics[width= 0.95\columnwidth]{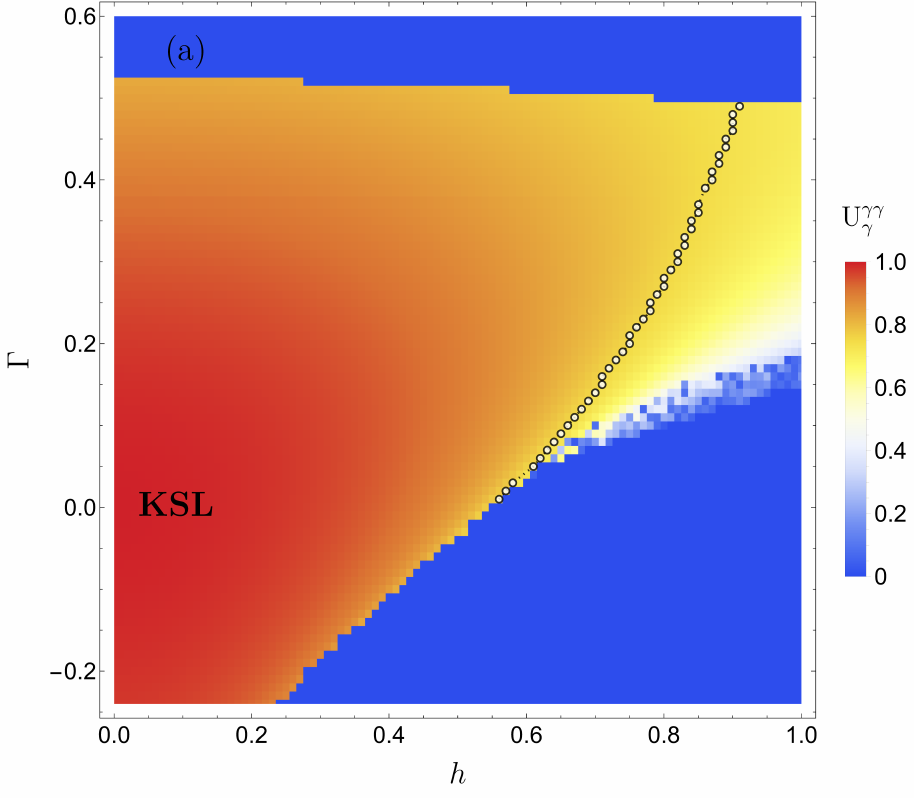}
\includegraphics[width= 0.95\columnwidth]{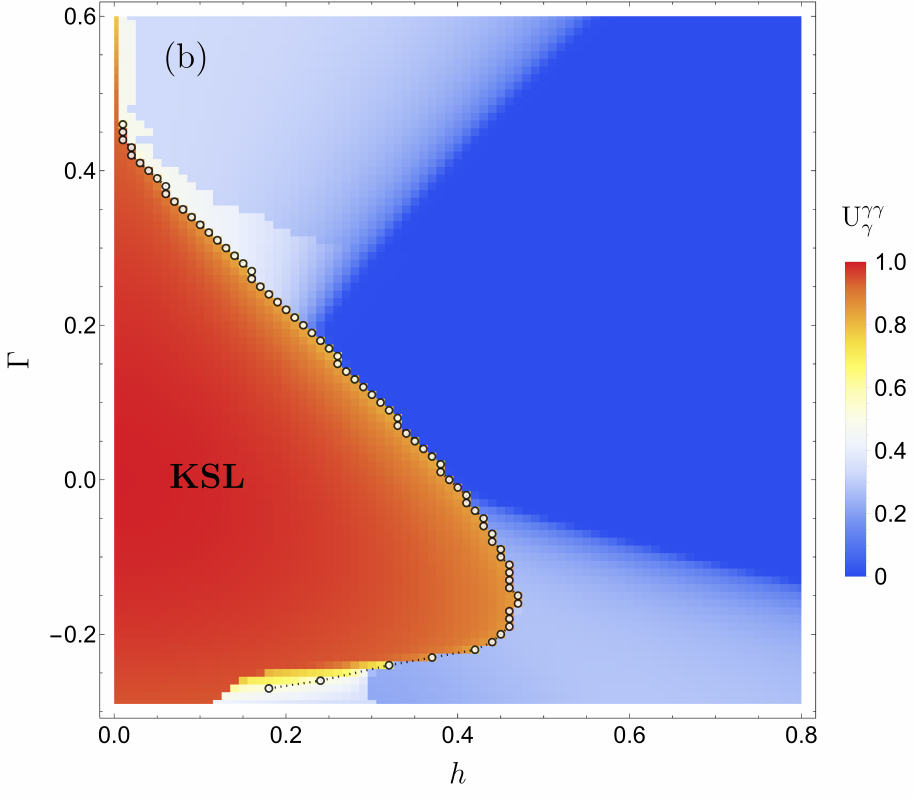}
\end{center}
\caption{ Mean-field parameter $\rmU_\gamma^{\gamma\gamma}$ for  fixed $K=-1$ and $J=0$ as a function of the $\Gamma$ interaction and the strength of the magnetic field along two directions: (a) $\mathbf h\parallel \hat{\mathbf c}$, perpendicular to the honeycomb plane; (b) an in-plane field $\mathbf h\parallel \hat{\mathbf a}$.   White circles represent critical points where the fermion gap closes at the $\Gamma$ point in the BZ.   The region labeled as KSL is identified with the Kitaev spin liquid phase.
}
\label{fig3}
\end{figure}

We further assess the stability of the  Kitaev spin liquid phase by evaluating the $\mathbb Z_2$ flux parameter.  In a homogeneous ground state, we have $W_p=(\rmU_{\gamma}^{\gamma\gamma})^6$. The result for the extended Kitaev model with $J=0$ and $\Gamma,h\neq0$ is shown in Fig.~\ref{fig3} 
for a magnetic field along the $\hat{\mathbf c}$ direction and for an in-plane field along  the $\hat{\mathbf a}$ direction (perpendicular to the $z$ bonds).   As expected,  $\rmU_\gamma^{\gamma\gamma}$ decreases as we increase $h$ or $\Gamma$. The dots in this figure mark the transition   where the gap $\Delta_f$ vanishes continuously. Note that $\rmU_\gamma^{\gamma\gamma}$ varies smoothly across the continuous  transition for $\mathbf{h}\parallel \hat{\mathbf c}$ and $\Gamma>0$. 

The results in Figs.~\ref{fig2} and  \ref{fig3} allow us to determine the parameter regime where both $\rmU_\gamma^{\gamma\gamma}$ and $\Delta_f$ vary smoothly and take values comparable to those at the exactly solvable point. In this regime,  we expect the mean-field approach to yield qualitatively correct results for the charge response of the Kitaev spin liquid phase. By contrast, the regime of strong magnetic fields should be identified with the partially polarized phase, whereas the regime of large  $|\Gamma|$ or $|J|$ harbors   magnetically  ordered phases  \cite{Rau2014,Yadav2016,Gordon2019,Yilmaz2022}. Here we do not explore the various   phases of the extended Kitaev model, whose nature is not completely settled \cite{Rousochatzakis2023}. Nevertheless, our mean-field results reproduce qualitative features of phase diagrams reported in the literature. For instance, we find that adding $\Gamma>0$ increases the critical magnetic field along the $\hat{\mb c}$ direction, but the Kitaev spin liquid phase shrinks as we tilt the field towards the plane,  in  agreement with exact diagonalization results \cite{Gordon2019}.  However,  in general the mean-field approach overestimates the value of the critical magnetic field  for a ferromagnetic Kitaev coupling in comparison with more accurate numerical methods \cite{Zhu2018,Hickey2019,Gordon2019,Kim2020}.  
 
\subsection*{Vortex charge density profile}

Inhomogeneous spin correlations can bring on a charge redistribution in Mott insulators  \cite{Bulaevskii_2008,Khomskii_2010}. We here discuss the charge density profile induced by the presence of $\mathbb Z_2$ vortices in a Kitaev spin liquid. 
In the Methods section, we derive the effective  charge imbalance operator in terms of two-spin operators and show how to compute its expectation value $\langle \delta n_l\rangle$   at lattice site $l$ using the Majorana mean-field approach. 

\begin{figure}[t]
\begin{center}
\includegraphics[width= .9\columnwidth]{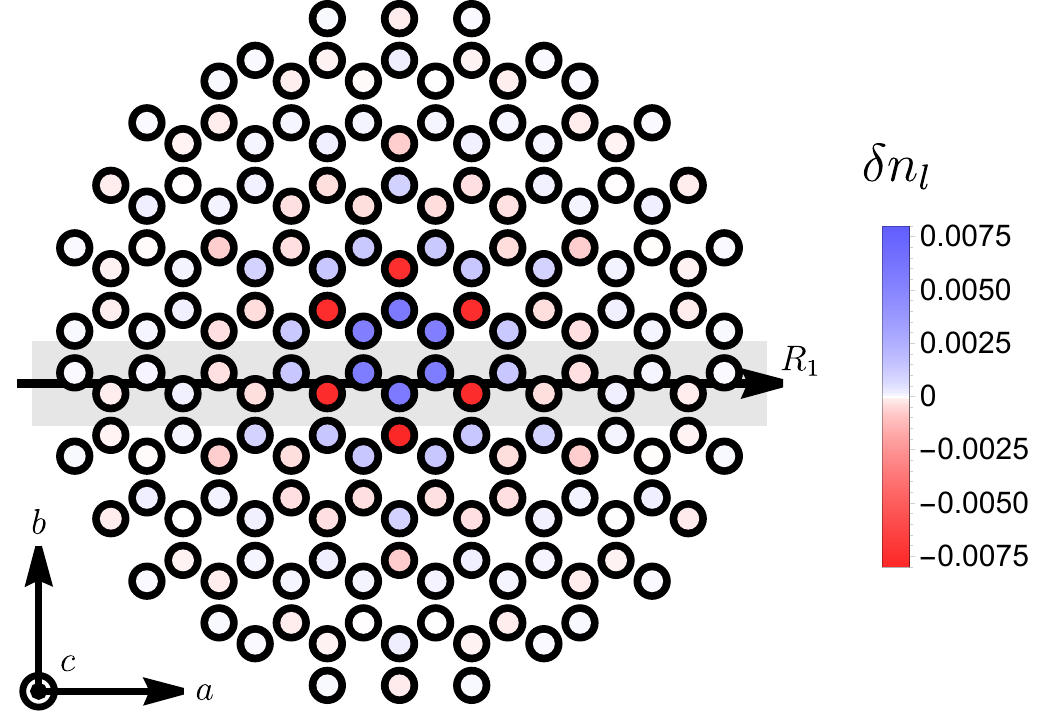}\\
\end{center}
\caption{Charge imbalance $\langle \delta n_l\rangle$ in a state with a vortex located in the central hexagon. As parameters of the Hubbard-Kanamori model, we use $t_1=13$ meV, $ t_2=160$ meV, $t_3=-33$ meV, $t_2^\prime=-60$ meV, $U = 2.6$ eV, and $J_H = 300$ meV. The values of $\delta n_l$ are in units of $|t_2^2 t_2^\prime/U^3| \approx 8.739 \times 10^{-5}$. The ratio between the exchange couplings calculated using Eq.~(\ref{eq:J}) are $\Gamma/|K|=0.20$ and $J/|K|=-0.02$.
We set the magnetic field $\mathbf{h}/|K|=0.2\hat{\mb c}$. The solid line marks the zigzag path considered in Fig.~\ref{fig5}.  
} 
\label{fig4}
\end{figure}

\begin{figure*}[t]
\begin{center}
\includegraphics[width=\textwidth]{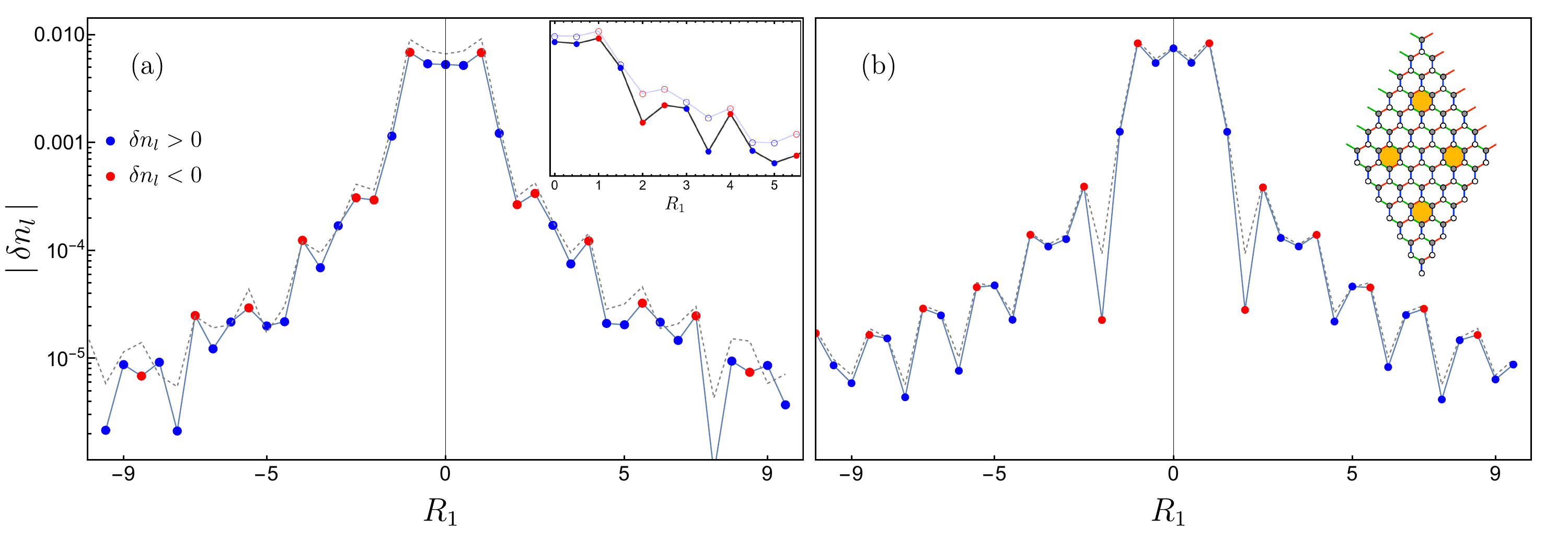}
\end{center}
\caption{Magnitude of the charge imbalance as a function of the position $R_1$ along the zigzag path represented by the black line in Fig.~\ref{fig4}.  The dots (connected by solid lines to guide the eye) correspond to the extended Kitaev model with exchange couplings   $\Gamma/|K|=0.2$  and $J/|K|=-0.02$.  Blue and red represent positive and negative charges, respectively. We set  $|\mathbf{h}|=0.3|K|$ and consider two field directions: (a) $\mathbf{h}\parallel \hat{\mb c}$, and (b) $\mathbf h\parallel \hat{\mathbf{z}}$.   For comparison, dashed lines represent the corresponding mean-field results for $\Gamma=0$ and otherwise identical parameters.  \new{The inset in (a) shows the corresponding case with $\Gamma/|K|=0.35$ and $J/|K|=-0.05$ for $\mathbf{h}\parallel \hat{\mb c}$ (filled circles), comparing with the results for $\Gamma/|K|=0.2$  and $J/|K|=-0.02$ in the main plot (empty circles).  The values of $\delta n_l$ are in units of $|t_2^2 t_2^\prime/U^3|$.} The inset in (b) shows the geometry with four equally spaced vortices on the torus with a smaller system size.  
}  
\label{fig5}
\end{figure*}

We consider an inhomogeneous state in which translation symmetry is broken by the presence of vortices. In this case, we analyze the mean-field Hamiltonian on a finite system with linear size $L$  along the directions of the primitive lattice vectors $\hat{\mb e}_1=\frac12\hat{\mb a}+\frac{\sqrt3}2\hat{\mb b}$ and $\hat{\mb e}_2=-\frac12\hat{\mb a}+\frac{\sqrt3}2\hat{\mb b}$, imposing periodic boundary conditions. To create vortices, we initialize the mean-field parameters in a configuration where we flip the sign of   $\rmU_{ij}^{\mu\nu}$ on bonds crossed by open strings. In the pure Kitaev model, this procedure generates exact eigenstates with two localized vortices at the ends of the string. In the extended Kitaev model, vortices become mobile excitations with effective bandwidths governed by the integrability-breaking perturbations \cite{Zhang2021,Joy_2022}.  In fact, for sufficiently large values of these perturbations, near the border of the Kitaev spin liquid phase in Fig.~\ref{fig3}, we observe that the vortex positions vary as we iterate the self-consistency equations. When this happens,  the string length decreases and the vortices move closer to each other until they annihilate, and the mean-field solution converges to the vortex-free ground-state configuration. However, for  $|\Gamma|,|J|, |h|\ll |K|$ and well separated vortices, we find a self-consistent solution with (metastable) localized vortices which corresponds to a local energy minimum in this sector of the Hilbert space. These results seem consistent with the real-time dynamics described by  time-dependent mean-field theory,  which show that  only when the perturbations are strong enough do vortices become mobile as signaled by the time decay of the fermion Green's function \cite{Cookmeyer2023}.   \new{In reality,  the  lifetime of a vortex is limited by processes in which two vortices meet and annihilate \cite{Joy_2022}, and can become arbitrarily long at low  temperatures due to the low vortex density, see the Supplemental 
Material (SM) \cite{SM}.} Focusing on the regime of small perturbations, we can then compute static spin correlations near vortices using position-dependent mean-field parameters $\rmU^{\mu\nu}_{ij}$ and $\rmV^{\mu\nu}_{i}$. \new{We consider a configuration with four equally spaced vortices, see inset of Fig.~\ref{fig5}(b), which preserves rotational symmetries and minimizes finite-size effects as compared to a two-vortex configuration.} Unless stated otherwise, we use $L=40$, so the distance between vortices is 20 unit cells. \new{The charge imbalance near a vortex is then effectively the property of a single vortex and finite-size effects only appear in long-distance tails \cite{SM}.
}   

In Ref.~\cite{Pereira2020},  the charge imbalance profile in   the vicinity of a vortex was investigated within the exactly solvable Kitaev model \cite{Kitaev2006}\be
H_K= \sum_{\left< ij\right>_\gamma} K_\gamma  \sigma_i^\gamma \sigma_j^\gamma  - \sum_{\braket{ij}_\alpha \braket{jk}_\beta}\!\kappa\, \sigma_i^\alpha \sigma_j^\gamma \sigma_k^\beta,  \label{eq:pure_K}
\ee
setting $K_\gamma=K$ for isotropic Kitaev interactions. The three-spin interaction breaks time-reversal symmetry while preserving integrability. The coupling constant derived from perturbation theory in the magnetic field is \cite{Kitaev2006} 
\be
\kappa= 0.338  \; \frac{h_xh_yh_z}{\Delta_{2\rm v}^2},\label{kappa}
\ee 
where $\Delta_{2\rm v}\approx 0.263|K|$  \cite{Panigrahi_2023} is the energy gap for creating two adjacent vortices at zero magnetic field. The prefactor in Eq.~\eqref{kappa} was obtained by fitting the fermion gap $\Delta_f=6\sqrt3 \kappa$ at low fields; see the inset of Fig.~\ref{fig2}.  Our mean-field results for the extended Kitaev model confirm the qualitative behavior obtained for the exactly solvable model; see   Fig.~\ref{fig4}.   The charge imbalance oscillates between positive and negative values as we vary the distance from the center of the vortex, identified with the plaquette where $W_p<0$. Moreover, as shown in Fig.~\ref{fig5}, the magnitude of $\langle \delta n_l\rangle$ decays exponentially with the distance from the vortex. The comparison with the result for $\Gamma=0$ (dashed lines in Fig.~\ref{fig5}) reveals \new{ that weak $\Gamma$ and/or $J$ interactions have an only minor effect on the ideal charge imbalance profile found in the pure 
Kitaev limit \cite{Pereira2020}. However, changing the magnetic field direction away from 
the $\hat{\mathbf{c}}$ direction can induce more pronounced charge oscillations, 
cf.~Fig.~\ref{fig5}(b), and thus has a more substantial effect.  }
 The value of $|\delta n_l|$ on sites around the vortex is of the order of $10^{-6}$,  producing local electric fields near the detection limit of state-of-the-art atomic force microscopy \cite{Pereira2020,Wagner2019,Bian2021,Yin2021}. Importantly, here we use estimates for the hopping and interaction parameters for bulk $\alpha$-RuCl$_3$, but the charge fluctuations can be greatly enhanced if the on-site repulsion $U$ is screened in a monolayer by the interaction with a substrate.

\begin{figure*}[t]
\begin{center}
\includegraphics[width=\textwidth]{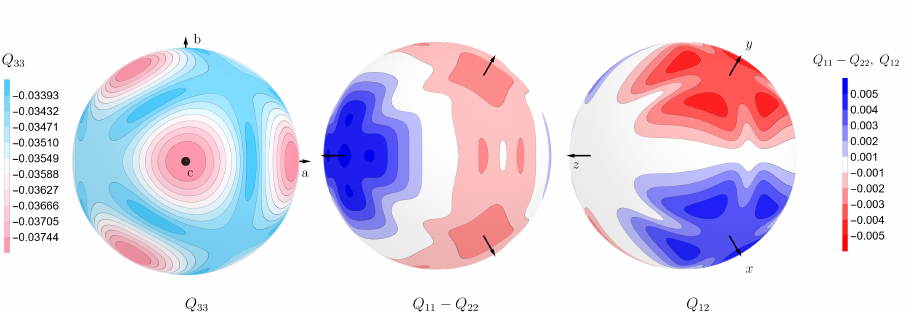}
\end{center}
\caption{Quadrupole components as a function of magnetic field direction, calculated using the exactly solvable  Hamiltonian in Eq.~\eqref{eq:pure_K}, i.e., for $J=\Gamma=0$. The coupling constants $K_\gamma$ and $\kappa$ were calculated using Eqs.~(\ref{kappa}) and (\ref{Kgamma}) with $|\mathbf{h}|=0.2 |K|$ and $\Delta_{2\rm v} = 0.263 |K|$.    
The scale is in units of $t_2^2t_2^\prime/U^3$ and we set the lattice spacing to unity.  Here we use $L=42$.
}
\label{fig6}
\end{figure*} 

Since the mean-field approach allows us to treat the Zeeman term nonperturbatively, we can go beyond the results of Ref.~\cite{Pereira2020} and analyze the dependence of the charge redistribution on the field direction. 
For a field along the $\hat{\mb c}$ direction, the charge imbalance profile is isotropic around the position of the vortex, up to small variations due to the finite distance between vortices in the finite-size system. As we tilt the magnetic field on the $ac$ plane (perpendicular to the $z$ bonds), a small anisotropy develops in a way that the charge imbalance is enhanced in the direction perpendicular to the field. This effect can be seen in Fig.~\ref{fig5}(b) as the difference between $\langle \delta n_l\rangle$ for the sites that belong to the hexagon that contains the vortex (three blue dots in the center, cf.~Fig.~\ref{fig4}).

We next quantify the anisotropy in the charge distribution by computing the electric multipole moments. \new{We note that the electric quadrupole moment has also been studied in the context of the spin nematic transition in the vortex-free ground state of a perturbed Kitaev model \cite{Yamada2021}.} In the limit of very large distance between vortices, the electric dipole moment vanishes because the system is invariant under spatial inversion about the vortex center.  The first nontrivial multipole moment is the traceless quadrupole tensor, with components  
 \be
Q_{\alpha\beta}=\sum_{l}\langle \delta n_l\rangle (3R_{l\alpha}R_{l\beta}-|\mb R_l|^2\delta_{\alpha\beta}).
\ee 
Here $\alpha,\beta\in \{1,2,3\}$ and $\mb R_l= R_{l1}\hat{\mb a}+R_{l2}\hat{\mb b}$ (with $R_{l3}=0$) is the position of site $l$, setting the lattice spacing to unity.  
Due to the finite system size, we calculate the quadrupole moment by summing over all sites within a finite radius around the vortex. This radius is taken to be slightly smaller than half the distance between vortices, but due to the exponential decay of $\langle \delta n_l\rangle$ with the distance from the vortex center, changing this radius causes only exponentially small changes in the quadrupole tensor. 
For a magnetic field along the $\hat{\mb c}$ direction, the rotational symmetry implies that the quadrupole tensor is diagonal and  $Q_{11}=Q_{22}=-Q_{33}/2$. As we vary the field direction, the anisotropy is manifested in the difference between $Q_{11}$ and $Q_{22}$ and in the off-diagonal element  $Q_{12}$. Note that $Q_{13}$ vanishes identically because $R_{l3}=0$.
  
In a first approximation, let us discuss the dependence of the quadrupole tensor on the magnetic field direction by treating the field  perturbatively in the pure Kitaev model.  For magnetic field directions not perpendicular to the lattice plane, the (often discarded) contribution from second-order perturbation theory generates an anisotropic renormalization of the Kitaev couplings. This effect is captured by the Hamiltonian in Eq.~(\ref{eq:pure_K}) with \be
K_\gamma=K-   \frac{(h_\gamma)^2}{\Delta_{2\rm v}}.\label{Kgamma}
\ee  
In Fig.~\ref{fig6}, we show the angular dependence of the quadrupole components  $Q_{33}$, $Q_{11}-Q_{22}$ and $Q_{12}$ calculated from the spin correlations for the model in Eq.~(\ref{eq:pure_K}). 
The component $Q_{33}$ does not change sign, but varies slightly around an average value with an angular dependence qualitatively similar to  $|h_xh_yh_z|$. In particular, $Q_{33}$ is maximum for a field along the $\hat{\mb c}$ direction, which may be interesting to maximize the intrinsic electric field produced at positions right above the vortex.    
On the other hand, the difference $Q_{11}-Q_{22}$ vanishes for $\mb h\parallel \hat{\mb c}$, but is maximum when the field points along the $\hat {\mb z}$ axis; this is the direction in which the anisotropy in the effective Kitaev couplings is maximized, with $K_z<K_x=K_y$.   Finally, $Q_{12}$ vanishes if we tilt the field along the high-symmetry $ac$ plane, but becomes nonzero for more general field directions.      

\begin{figure}[t]
\begin{center}
\includegraphics[width=\columnwidth]{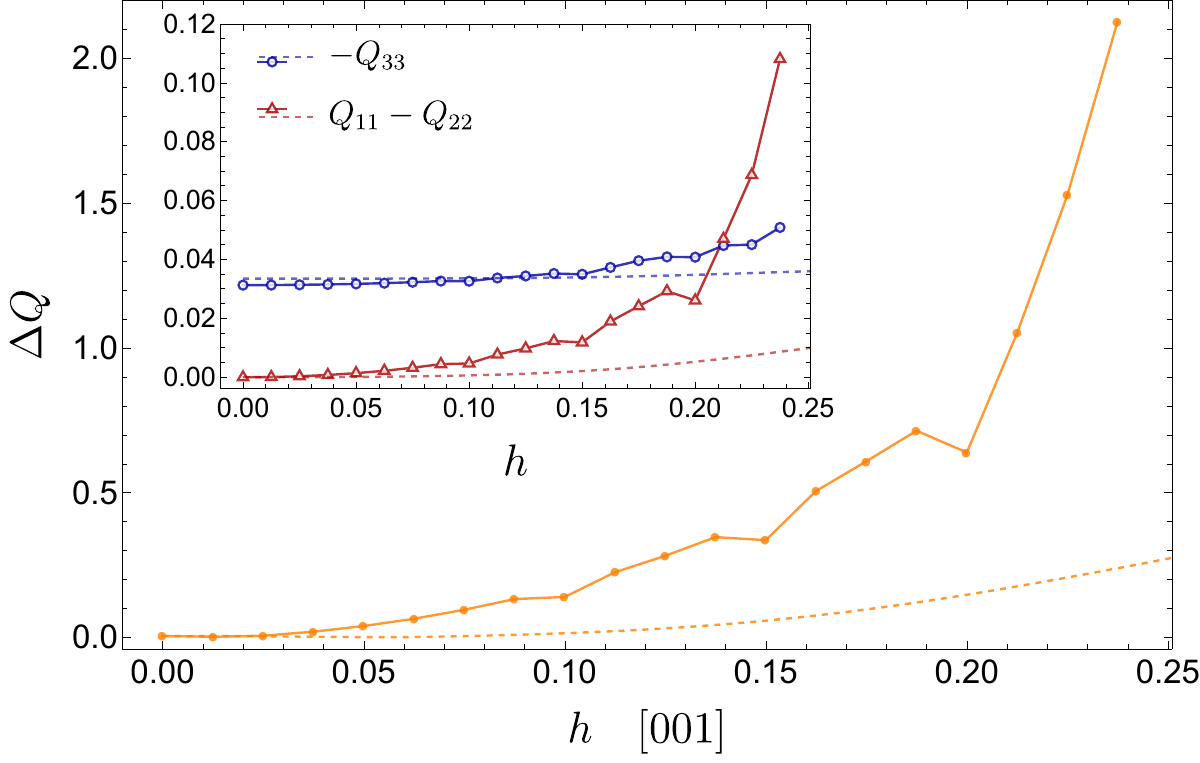}
\end{center}
\caption{Dependence of the quadrupole anisotropy $\Delta Q= (Q_{11}-Q_{22})/|Q_{33}|$ on the magnetic field strength for $\mb h=h\hat{\mathbf{z}}$. Dashed lines follow from the solvable Hamiltonian in Eq.~\eqref{eq:pure_K} with $K=-1$. 
Symbols represent the corresponding mean-field results for the extended Kitaev model with $\Gamma/|K| = 0.3$ and $J=0$.   
Inset: quadrupole components $Q_{33}$ and $Q_{11}-Q_{22}$ (in units of $|t_2^2t_2^\prime/U^3|$ and setting the lattice spacing to unity).  
} 
\label{fig7}
\end{figure}

The spin correlations calculated within the mean-field approach for the extended Kitaev model lead to the same qualitative dependence on the field direction as in Fig.~\ref{fig6}.  
To maximize the anisotropy in the quadrupole tensor, we focus on the direction $\mb h=h\hat{\mb z}$, in which case all off-diagonal components vanish, and analyze how the diagonal components vary with the strength of the magnetic field. Here  it is convenient to introduce the dimensionless anisotropy parameter $\Delta Q={(Q_{11}-Q_{22})}/|Q_{33}|$. As shown in Fig.~\ref{fig7}, $\Delta Q$ increases with $h$, and the effect is more pronounced in the presence of the $\Gamma$ interaction. 
\new{We have also studied the case $\Gamma<0$ and find qualitatively similar results \cite{SM}.}

\begin{figure}[t]
\begin{center}
\includegraphics[width=\columnwidth]{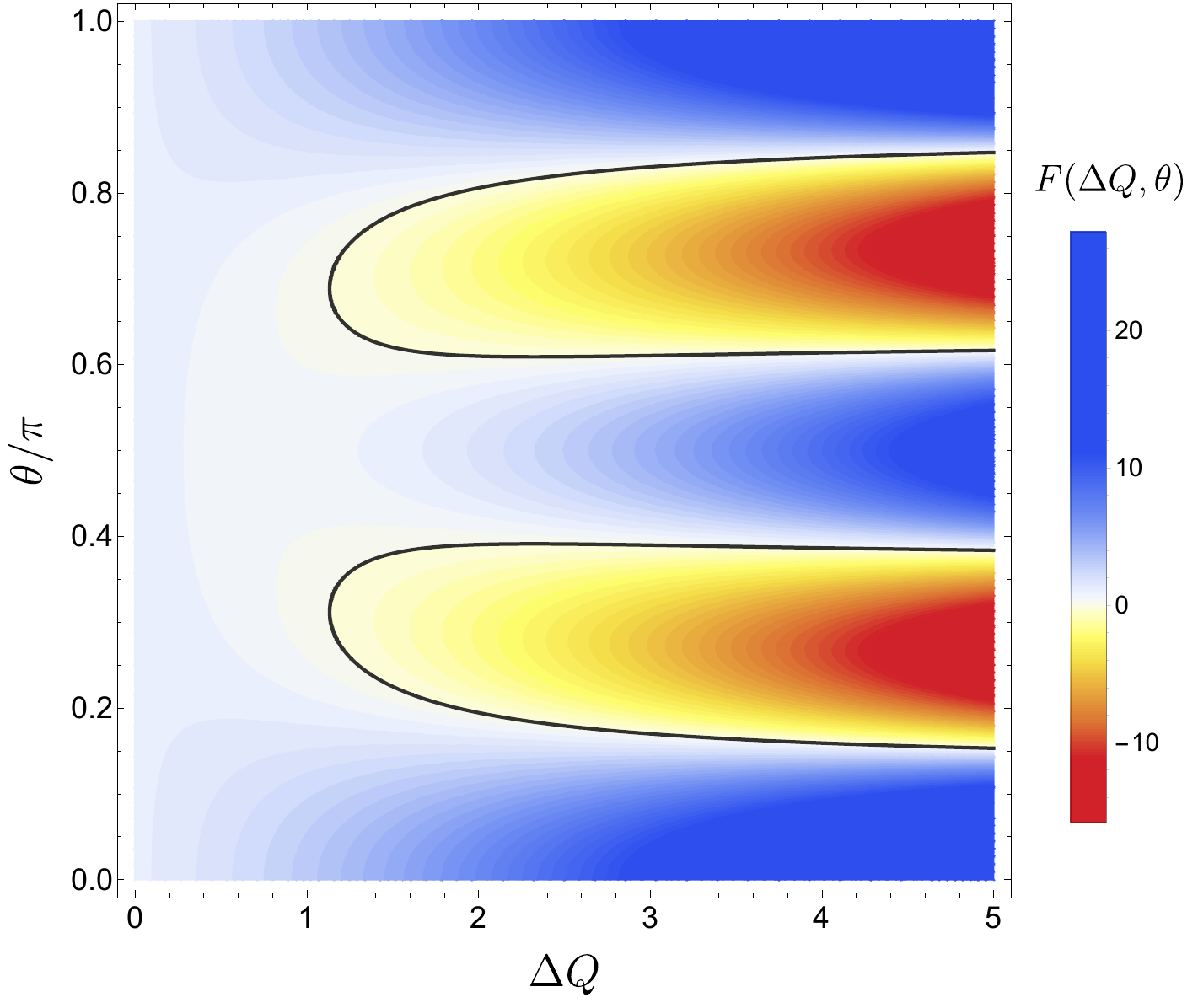}
\end{center}
\caption{  Function $F(\Delta Q,\theta)$ that governs the sign of the quadrupolar interaction for $\Delta Q>0$. For $\Delta Q<0$, see the relation below   Eq.~\eqref{eq:E-cos}.  The dashed line marks the critical value $|\Delta Q| = \sqrt{9/7}$, below which the interaction is always repulsive. The black solid line corresponds to $F(\Delta Q,\theta)=0$.   }
\label{fig8}
\end{figure}

The spatial anisotropy of the charge density profile affects the electric quadrupole interaction between vortices. Suppose the first vortex is located at the origin and the second one at   $\mb r=x_1\hat{\mb a}+x_2\hat{\mb b}$, with $r=|\mb r|$ much larger than the length scale in the decay of $\delta n_l$.  The interaction is given by the 
energy $\mathcal{E}$ of the quadrupole tensor $Q^{(2)}$  of the second vortex in the electrostatic potential $V^{(1)}$ generated by  the first vortex,
\begin{equation}
\begin{split}
\mathcal{E}&= \frac{1}{6}  \sum_{\alpha\beta} Q^{(2)}_{\alpha \beta}  \,  \partial_\alpha \partial_\beta V^{(1)}(\mb r)  ,
\end{split}
\end{equation}
where $V^{(1)}(\mathbf{r}) =  \frac{1}{2r^5}  \sum_{\alpha\beta} x_\alpha x_\beta Q_{\alpha \beta}^{(1)}$. 
Since well-separated vortices generate the same charge distribution, we now assume $Q^{(1)}=Q^{(2)}=Q$. 
As a result, the quadrupolar interaction can be written as
\begin{align}
    \mc E&=\frac{1}{12}\Big[ 
    \frac{35}{r^9}\left(\mathbf{r}\cdot{Q}\!\cdot\mathbf{r}\right)^2\!\!
    -\!\frac{20}{r^7}\left(\mathbf{r}\cdot{Q}^2\!\cdot\mathbf{r}\right)\!
    +\!\frac{2}{r^5}\mathrm{ tr}\left( {Q}^2\right)\!\Big].
\end{align}
When the magnetic field varies along the $ac$ plane, the quadrupole tensor is diagonal and we obtain
$\mc E = \frac{Q^2_{33}}{ r^5} F(\Delta Q, \theta)$.
Here $\theta$ is the angle between $\mb r$ and $\hat{\mb a}$, and we use 
\bea
F(\Delta Q,\theta)&=& 
    \frac{9}{8}
    +\frac{5}{4} \cos (2\theta)    \Delta  Q \nonumber\\
    &&
    +\left[ \frac{35}{24} \cos^2 (2\theta)-\frac{2}{3} \right]\Delta Q^2
 \label{eq:E-cos},
\eea
with the property $F(-\Delta Q,\theta)=F(\Delta Q,\pi/2-\theta)$. 
 In particular, $\Delta Q=0$ for a magnetic field along the $\hat {\mb c}$ direction; in this case,  the quadrupolar interaction becomes strictly repulsive and independent of $\theta$.  However,
 as illustrated in Fig.~\ref{fig8}, the interaction can change sign for some particular directions of $\mb r$ if the anisotropy is strong enough. The attractive regime appears for $|\Delta Q| > \sqrt{9/7}\approx 1.13$.   According to the result in Fig.~\ref{fig7}, this regime becomes accessible for sufficiently large $h$ and $\Gamma$ with $\mb h$ along the $\hat{\mb z}$ direction. We note that already in the pure Kitaev model, vortices have an effective interaction that depends on the vortex separation  \cite{Lahtinen2012}.  The charge redistribution discussed here provides a mechanism to make this interaction spatially anisotropic. In the extended Kitaev model,  where vortices acquire a small mobility \cite{Joy_2022}, the charge density profile must be carried along with the slow vortex motion, 
 and the anisotropic interaction may cause some nontrivial dynamics in a system of dilute vortices. \new{Importantly, the quadrupole interaction decays algebraically with the distance between vortices; thus, at large distances it dominates over other sources of vortex-vortex interactions that are expected to decay exponentially \cite{Lahtinen2012}. }

\subsection*{Electrical manipulation of vortices}

\begin{figure}[t]
\begin{center}
\includegraphics[width=\columnwidth]{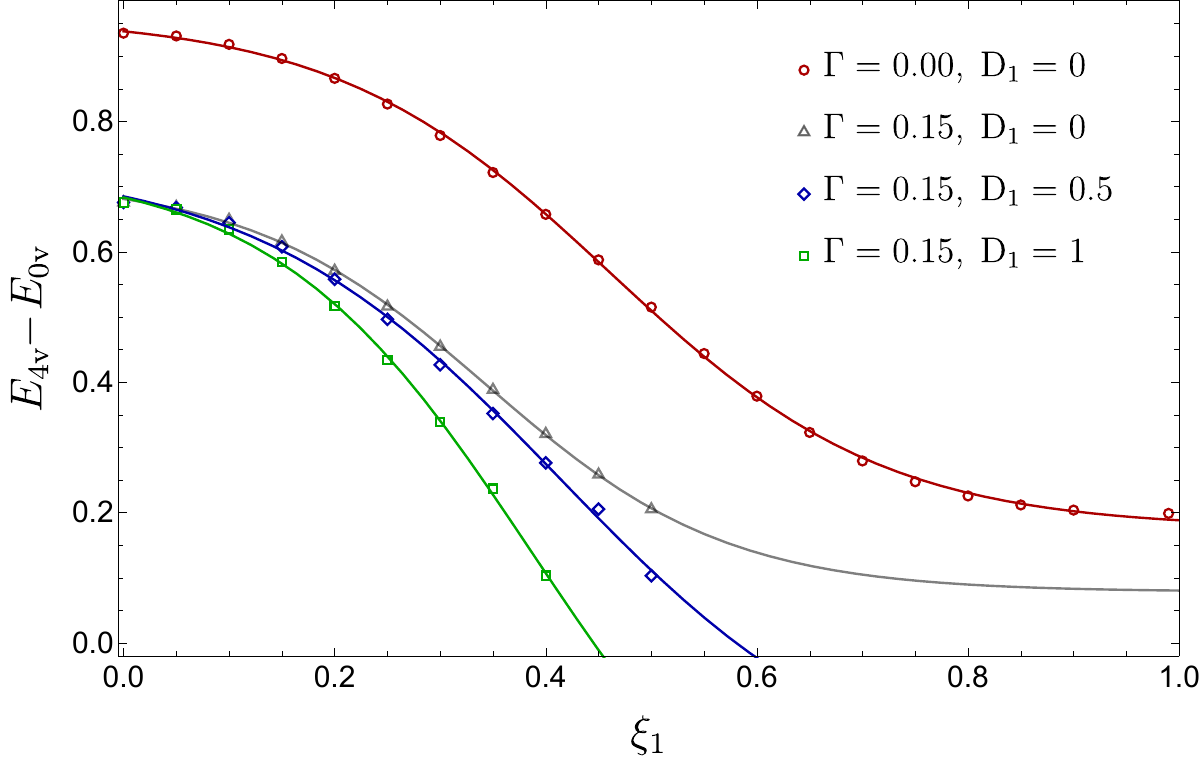}
\end{center}
\caption{Energy of the four-vortex state vs applied electrostatic potential $V_0$, with the dimensionless quantity
$\xi_1= eV_0/(U-3J_H)$, for different values of $\Gamma$ and of the DM coupling $D_1$; see
Eq.~(\ref{eq:ev-DM}). Symbols represent mean-field results for the extended Kitaev model with $K=-1$,  $J=0$ and \new{the magnetic field} $\mathbf{h}=0.2\,\hat{\mathbf{c}}$.  The linear system size is $L=28$, and solid lines are a guide to the eye only.
They were obtained by a fit to
the function $a+b \tanh[c(\xi_1-d)]$.
}
\label{fig9}
\end{figure}

We now consider the effect of a local electrostatic potential on vortices. Going back to the Hubbard-Kanamori model,  we couple the hole density to a potential $V_0$ on the six sites surrounding a hexagonal plaquette $p$ where a vortex is located. This local potential can be generated by the electric field of a scanning tunneling microscope (STM) tip.  
 Redoing the derivation of the effective spin Hamiltonian by second-order perturbation theory, we find that the local potential modifies the couplings on bonds between sites in $p$ and their nearest neighbors outside $p$; see Eq.~\eqref{eq:ev-J}.
In addition, the local electric potential  breaks inversion symmetry and  generates a Dzyaloshinskii-Moriya (DM) interaction \cite{Jang2021}. Microscopically, the DM interaction stems from crystal field splittings  in the atomic Hamiltonian and asymmetries in the hopping matrix due to lattice distortions \cite{Winter2017rev}. We investigate this effect phenomenologically by adding  to the effective spin Hamiltonian \eqref{Hwithfield} the term
\begin{equation}
   H_{\text{DM}} = \sum_\gamma\! \sum_{\braket{ij}_{\gamma} }     \; D_{ij}  \, \left( \sigma_i^\alpha \sigma_j^\beta -\sigma_i^\beta \sigma_j^\alpha    \right),
\end{equation}
where $(\alpha\beta\gamma)$ is a cyclic permutation of $(xyz)$. The coupling $D_{ij}=D(V_0)$ is taken to be independent of the bond type $\gamma$ but restricted to the bonds exterior to the plaquette with the 
local potential. For the DM coupling, we assume  \new{\cite{SM}}
\be \label{eq:ev-DM}  
    D(V_0) =  \xi_1 D_1 |K(0)|, \quad \xi_1=\frac{eV_0}{U-3J_H},
\ee
such that $D(V_0)\propto V_0$ with a dimensionless free parameter $D_1$.  
In fact, for $V_0=0$, the DM coupling is absent since it will be generated by the tip potential.

In the solvable Kitaev model,  the local electric potential  lowers the energy of an isolated vortex  with respect to the vortex-free configuration, but never closes the vortex gap \new{in the absence of the DM interaction} \cite{Pereira2020}. In that case, this effect can be used to attract and bind vortices that have been created by some other mechanism, such as thermal fluctuations, but it does not induce vortices in the ground state of the system.  
Using the mean-field approach, we can now analyze how the vortex gap varies with the electric potential in the extended Kitaev model. We consider again the configuration with four  equally spaced vortices, see the inset in Fig.~\ref{fig5}(b), and apply   the electric potential on  the four corresponding plaquettes.   The difference between the energy  $E_{4\rm v}$   of  this four-vortex configuration and the energy $E_{0\rm v}$  of the vortex-free state is equal to four times the vison gap. As shown in Fig.~\ref{fig9}, the vison gap monotonically decreases with the applied electric potential, and it is further reduced for nonzero $\Gamma$ and
finite DM coupling $D_1$. 
When the gap becomes too small, we encounter difficulties  in the convergence of the mean-field equations. However, the extrapolation of the results indicates that the  gap vanishes for sufficiently large $V_0$. As a consequence, we predict that it is 
possible to create (or remove) vortices by modulating the local interactions, in agreement with the results of Ref.~\cite{Jang2021}.  We emphasize that this remarkable functionality arises due to the interplay between $\Gamma$ interactions and the local DM terms induced by an STM tip. 
\new{From Fig.~\ref{fig9}, we observe that $\xi_1\sim 0.5$ is sufficient to create vortices. 
Using the parameters listed in Fig.~\ref{fig4}, we find that this corresponds to realistic 
tip voltages of the order of $V_0< 1$~V. } 

\section*{Methods}
\subsection*{Extended Kitaev model}

The $JK\Gamma$ model in Eq.~\eqref{Hwithfield} follows by projecting the 
three-orbital Hubbard-Kanamori Hamiltonian on the honeycomb lattice, $H_{\rm HK} =  V +H_{\rm so}+T$,
to the low-energy sector spanned by a single hole per site. 
On-site interactions are encoded by 
\begin{equation}
    V=\sum_i \left[ \frac{U-3J_H}{2}(\bar{N}_i-1)^2     \!-\!2J_H \mathbf{S}_i^2 
    \!-\!\frac{J_H}{2} \mathbf{L}_i^2 \right], 
\end{equation} 
where $U$ is the repulsive interaction strength, $J_H$ is Hund's coupling, and  the operators $\bar N_i$, $\mathbf S_i$ and $\mathbf L_i$ are the total number,  spin and orbital angular momentum of holes at site $i$.
The operator $h^\dagger_{i\alpha \sigma} $ creates a hole at site $i$ with spin $\sigma \in \{ \uparrow, \downarrow \}$ and orbital 
 $\alpha \in\{ x,y,z\}$  for  $yz$, $xz$, and $xy$ orbitals, respectively.
Defining the spinor $h_i^\dagger=(h^\dagger_{ix\uparrow},h^\dagger_{iy\uparrow},h^\dagger_{iz\uparrow},h^\dagger_{ix\downarrow},h^\dagger_{iy\downarrow},h^\dagger_{iz\downarrow})$, we   write
\be
\bar N_i=h^\dagger_ih^{\phantom\dagger}_i,\quad \mathbf S_i=\frac12h^\dagger_i(\boldsymbol \sigma\otimes \mathbb 1_3)h^{\phantom\dagger}_i,\quad \mb L_i=h^\dagger_i(\mathbb 1_2 \otimes \mb l )h^{\phantom\dagger}_i,
\ee
where $\boldsymbol \sigma$ is the vector of Pauli matrices acting in spin space and $\mb l=(l_x,l_y,l_z)$ is a vector of $3\times3$ matrices that represent the effective $l=1$ angular momentum of the $t_{2g}$ states \cite{Pereira2020}. The  spin-orbit coupling term
$H_{\rm so}=\lambda \sum_{i\alpha}  h^\dagger_i (\sigma^\alpha\otimes l^\alpha)h^{\phantom\dagger}_i$
splits the degeneracy of the $t_{2g}$ manifold. At each site, the low-energy subspace is  spanned by the states
\bea
\left|+\right\rangle&=&\frac1{\sqrt3}\left(-\left|z,\uparrow\right\rangle-i\left|y,\downarrow\right\rangle-\left|x,\downarrow\right\rangle\right),\nonumber\\
\left|-\right\rangle&=&\frac1{\sqrt3}\left(\left|z,\downarrow\right\rangle+i\left|y,\uparrow\right\rangle-\left|x,\uparrow\right\rangle\right),\label{jeff12}
\eea
which are associated with  total angular momentum $j_{\rm eff}=\frac12$. Finally, 
the hopping term in $H_{\rm K}$ has the form 
$T = - \sum_{ij} h_i^\dagger  \left( \mathbb{1}_2 \otimes \mathbf{T}_{ij } \right) h_j^{\phantom{\dagger}}.$
The hopping matrix $\mb T_{ij}$  in orbital space depends on the orientation of the bond between sites $i$ and $j$. We label  the bonds on the honeycomb lattice by $\gamma\in \{x,y,z\}\equiv\{1,2,3\}$ 
corresponding to nearest-neighbor vectors $\boldsymbol\delta_x=\frac{1}{2}\hat{\mb a}+\frac1{2\sqrt3}\hat{\mb b}$,  $\boldsymbol\delta_y=-\frac{1}{2}\hat{\mb a}+\frac1{2\sqrt3}\hat{\mb b}$, and  $\boldsymbol\delta_z=-\frac1{\sqrt3}\hat{\mb b}$, respectively. We parametrize the hopping matrix for a nearest-neighbor $z$ bond as \cite{Rau2014}
$
    \mathbf{T}_{\braket{ij}_z} =
    \left( 
    \begin{array}{ccc}
         t_1 & t_2  & {t_4} \\
         t_2 & t_1  & {t_4} \\
         {t_4} & {t_4}  & t_3 \\
    \end{array}
    \right)  .
$
The hopping matrix for $x$ and $y$ bonds follows by cyclic permutation of the orbital indices.  Microscopically, the hopping parameters   are associated with   direct hopping between $d$ orbitals or  hoppings mediated by the ligand ions. Neglecting trigonal distortions for simplicity, hereafter we  set  $t_4=0$   \cite{Winter2017rev,Rau2014}.  

The effective spin Hamiltonian for the Mott insulating phase can now be derived by applying perturbation theory in the regime $U,J_H\gg \lambda \gg t_1,t_2,t_3$. We use the canonical transformation
\bea
    \tilde{H}_{\rm HK} &=& e^{ S}   {H}_{\rm HK}   e^{-S} \nonumber\\
    &=& {H}_{\rm HK}  + [S,{H}_{\rm HK} ]+\frac{1}{2}[S,[S,{H}_{\rm HK} ]] + \cdots \, . \label{eq:II-1-S-transf}
\eea
The anti-Hermitian operator $S=\sum_{k=1}^{\infty}S_k$ is chosen so that $S_k$ eliminates the terms that change the hole occupation numbers  $\bar N_i$ at $k$-th order in the hopping parameters.  We can write $S_k=S_k^+-S_k^-$, where $S_k^+$ creates   excitations with $\bar N_i\neq 1$ and $S_k^-=(S_k^+)^\dagger$. For the calculation of the effective spin Hamiltonian, it suffices to consider the first-order term $S_1 = S_1^{+} - S_1^{-}$, with 
\begin{equation}
    S_1^{+}  = \sum_{ij}\sum_{\ell=0}^2 \frac{1}{\Delta E_{\ell}} \mathcal{P}^{(2)}_{i,\ell} \, h_i^\dagger ( \mathbb 1\otimes {\mathbf{T}}_{ij} )h_j^{\phantom{\dagger}} \, \mathcal{P}^{(1)}_{j}.  
\end{equation}
Here $\mathcal{P}^{(1)}_{j}$ is a projector onto the subspace of a single hole at site $j$ and $ \mathcal{P}^{(2)}_{j,\ell}$ projects onto the subspace of two holes with total angular momentum $\ell\in\{0,1,2\}$. The  excited states have energies $\Delta E_{\ell}$ given by  
\be \label{deltaEl}
\Delta E_0 = U + 2 J_H,\quad \Delta E_1=U-3J_H, \quad \Delta E_2=U-J_H,
\ee 
with $J_H<U/3$ in the Mott insulating phase. We then take $H=\mc P_{\rm low}\tilde H_{\rm HK}\mc P_{\rm low}$,
where $\mc P_{\rm low}=\prod_i \left(\ket{+_i}\bra{+_i}+ \ket{-_i}\bra{-_i}\right)$
is the projector onto the low-energy  subspace restricted to
$j_{\text{eff}}=\frac12$ states at every site. We thereby arrive 
at the $JK\Gamma$ model \cite{Rau2014}, 
\begin{equation}
        H\!=\sum_{\left< ij\right>_\gamma\!\!}\!\Big[ 
        J  \bm{\sigma}_i \cdot \bm{\sigma}_j 
        + K  \sigma_i^\gamma \sigma_j^\gamma  
        + \Gamma  \big( \sigma_i^\alpha \sigma_j^\beta+\sigma_i^\beta \sigma_j^\alpha \big)     \Big]  , \label{eq:H_ext}
\end{equation} 
with an implicit sum over bond type $\gamma$, and $\alpha,\beta$ chosen so that $(\alpha\beta\gamma)$ is a cyclic permutation of $(xyz)$.  The couplings are
\begin{eqnarray}
        J   &= & \frac{1}{27} \left[
    \frac{(2  t_{1}+ t_{3})^2}{ \Delta E_0}\!+\frac{6t_{1} ( t_{1}+2  t_{3})}{\Delta E_1} 
    +\frac{2(t_{1}-t_{3})^2}{\Delta E_2}  \right],  \nonumber
 \\     K&=&\frac{2J_H}{9} \frac{ (t_{1}-t_3)^2-3  t_{2}^2   }{\Delta E_1\Delta E_2}  , \quad
     \Gamma =  \frac{4J_H}{9}  \frac{t_{2} ( t_{1}- t_{3})}{\Delta E_1 \Delta E_2}  .\label{eq:J}
\end{eqnarray}
In the limit $t_1,t_3\to 0$ and $t_2\neq0$,  Eq.~(\ref{eq:H_ext}) reduces  to the exactly solvable Kitaev model \cite{Kitaev2006} with a ferromagnetic Kitaev interaction ($K<0$).
Finally, in the presence of a potential $V_0$, the respective couplings are renormalized according to 
\bea
    J(V_0) &=& \frac{1}{27} \left[
    \frac{(2  t_{1}+ t_{3})^2}{ (1-\xi_0^2)\Delta E_0} +\frac{6t_{1} ( t_{1}+2  t_{3})}{(1-\xi_1^2)\Delta E_1}
    \right.\nonumber\\
   &&\left.
     + \frac{2(t_{1}-t_{3})^2}{(1-\xi_2^2)\Delta E_2} \right], \label{eq:ev-J}´  
    \\ \nonumber 
   \frac{K(V_0)}{K(0)} &=& \frac{\Gamma(V_0)}{\Gamma(0)} =  
  \frac{1+\xi_1 \xi_2}{(1-\xi_1^2)(1-\xi_2^2)},
\eea
where $\xi_\ell = eV_0/\Delta E_\ell$.

\subsection*{Mean-field Hamiltonian}

Using the mean-field parameters in Eq.~\eqref{eq:MF_parameters}, the Majorana mean-field Hamiltonian for Eq.~\eqref{eq:H} is given by  
\begin{equation}
    H_{\mathrm{MF}}=\sum_{ {ij} }  \frac{\im}{4} c_i^{\T}   \rmA_{ij}  c_j +\sum_i  \frac{\im}{4} c_i^{\T}   \rmB_{i}  c_i-\rmC . \label{eq:HMF}
\end{equation}
The  first  term on the right-hand side couples Majorana fermions on nearest-neighbor  bonds $\braket{ij}_\gamma$ via the $4\times 4$ bond-dependent matrix     
\begin{equation}
    \rmA_{ij} = 2 \sum_{\alpha\beta}\mathbf{J}_{ij}^{\alpha\beta}  \rmN^\alpha \rmU_{ij} \rmN^\beta     .  \label{eq:A}
\end{equation}
The on-site term involves the matrix 
\begin{equation}
    \rmB_{i}= \sum_{j\in \mc V_i}   \sum_{\alpha\beta}\mathbf{J}_{ij}^{\alpha\beta} 
    \rmN^\alpha \tr\!\left( \rmV_{j}^\T \rmN^\beta \right) 
    +\sum_\gamma(\lambda_i^\gamma  \rmG^\gamma -    2h^\gamma  \rmN^\gamma )     
    , \label{eq:B}
\end{equation}
where $\mc V_i$ denotes the set of nearest neighbors of site $i$.  
Finally, the constant term is 
\begin{equation}
\begin{split}
       \rmC = \frac{1}{8}\sum_{ij}  \sum_{\alpha\beta}\mathbf{J}_{ij}^{\alpha\beta} \Big[ 
       &   \tr\!\left( \rmV_{i}^\T \rmN^\alpha \right)\tr\!\left( \rmV_{j}^\T \rmN^\beta \right) 
       \\ 
       + 2 \, & \tr\!\left( \rmU_{ij}^\T \rmN^\alpha   \rmU_{ij} \rmN^\beta \right)   \Big]   
    .  \label{eq:C}
\end{split}
\end{equation}
We diagonalize Eq.~\eqref{eq:HMF} for $N$ unit cells of the honeycomb lattice with periodic boundary conditions by using  
\begin{equation}
    c =\sqrt2\, \mathbb{U} \, \left( \begin{array}{c}  d^{\phantom{\dagger}} \\ d^\dagger \end{array} \right)  , \quad 
    \mathbb{U} = \left( \begin{array}{cc} \mathbb{U}_{<} & \mathbb{U}_{>} 
    \end{array}\right) ,
\end{equation}
where $c$ is a vector defined from  $8N$ Majorana fermions, $\mathbb{U}$ is a unitary transformation, and $d$ is a $4N$-component vector of annihilation operators of  complex fermions. The columns of $\mathbb{U}_{<(>)}$ correspond to the eigenvectors of the mean-field Hamiltonian with negative (positive) energy. The mean-field ground state is the state annihilated by all $d$ operators, from which we obtain the self-consistency conditions 
\begin{equation}
        \braket{ \,  \im   c_\I \,  c_\J  \, } = \im \big(\mathbb{U}_{<}^{\pd}  \mathbb{U}_{<}^\dagger  \big)_{\I\J}, \label{eq:MF_eq}
\end{equation}
where $\I=(i,\mu)$ and $\J=(j,\nu)$ combine site and fermion flavor indices. We obtain the mean-field parameters in Eq.~\eqref{eq:MF_parameters} by setting $i$ and $j$ to be either nearest neighbors or the same site. Together with the mean-field Hamiltonian, Eq.~(\ref{eq:MF_eq}) defines a set of self-consistent equations which we then solve numerically. 

In our approach, we require that the constraint in Eq.~(\ref{eq:G=0}) is satisfied 
by the mean-field solution as accurately as possible.  Since $ic^{\rm T}_i\rmG^\gamma c^{\phantom{T}}_i$ are linear combinations of  operators with eigenvalues $\pm1$, we define the quantities ${\cal G}_i^\gamma \equiv \frac{1}{4}
|\braket{c^{\rm T}_i\rmG^\gamma c^{\phantom T}_i}|$ for the mean-field ground state average,
with $0\le {\cal G}^\gamma_i\le 1$.  For zero magnetic field and in the absence of magnetic order, 
the constraints are automatically satisfied, ${\cal G}_i^\gamma=0$, since $\rmV^{0\gamma}_i=\rmV^{\alpha\beta}_i=0$. To describe the Kitaev spin liquid phase at finite magnetic field, we tune the Lagrange multipliers $\lambda_i^\gamma$ contained in $\rmB_i$ in order to minimize the violation of the constraint measured by ${\cal G}^\gamma_i$. For all results shown below, we guarantee 
${\cal G}_i^\gamma<0.05$  for all values of $(\gamma,i)$.  
In the homogeneous case (cf.\ Figs.~\ref{fig1}, \ref{fig2} and \ref{fig3}),
the largest violations occur in the vicinity of phase transitions. 
Away from transitions, we instead find 
${\cal G}_i^\gamma< 10^{-3}$. Similarly, in the presence of vortices, the 
largest violations occur near a vortex but they are always bounded as specified. 

\subsection*{Charge density coefficients}

\begin{figure}[t]
\begin{center}
\includegraphics[width= .7\columnwidth]{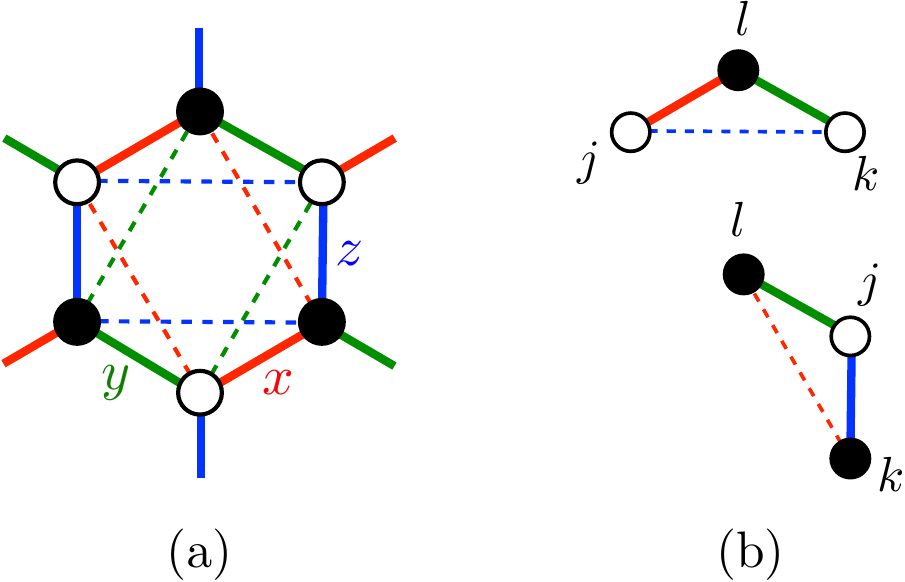}
\end{center}
\caption{Honeycomb lattice with nearest- and next-nearest-neighbor hopping. 
(a) Red, green and blue lines correspond to $\gamma=x,y,z$ bonds, respectively. 
The hopping matrix on nearest-neighbor bonds  (solid lines) is written in terms of hopping parameters $t_1$, $t_2$, $t_3$, and $t_4$. On second-neighbor bonds (dashed lines), we consider a single hopping parameter $t_2'$. 
(b) Examples of triangles that contribute to the effective charge density operator at site $l$.
}
\label{fig10}
\end{figure}

Consider the hole density operator $\bar N_l$ at site $l$ in the Hubbard-Kanamori model.  Using the  canonical transformation in  Eq.~(\ref{eq:II-1-S-transf}), we can write the effective charge imbalance operator in the low-energy sector as 
\be
\delta {n}_l=\mc P_{\rm low}e^S (\bar N_l -1)e^{-S} \mc P_{\rm low}.\label{transformdeltan}
\ee
We calculate $\delta  {n}_l$ using perturbation theory to leading order in the hopping matrix $\mb{T}_{ij}$. In systems with bond-inversion symmetry like the Hubbard-Kanamori model, the first non-vanishing contribution appears at third order and is associated with virtual processes in which  an electron or  hole moves around a triangle \cite{Pereira2020,Bulaevskii_2008,Khomskii_2010}. To obtain this leading contribution, we  generalize the hopping matrix to include hopping between next-nearest-neighbor sites on the honeycomb lattice. We denote by $\langle \langle ij\rangle\rangle_\gamma$ a second-neighbor bond perpendicular to nearest-neighbor  $\gamma$ bonds, see Fig.~\ref{fig10}(a). Sizeable second- and third-neighbor hopping parameters have been calculated for Kitaev materials using {\it ab initio} methods \cite{Winter_2016,Winter2017rev}. For simplicity,  we consider only the dominant second-neighbor hopping, which on $z$ bonds is described by the   matrix 
$\mathbf{T}_{{\braket{\braket{ij}}}_z}  =    \left( 
    \begin{array}{ccc}
         0 & t_2'  & 0 \\
         t_2' & 0  & 0 \\
         0 & 0  & 0 
    \end{array}
    \right)$.
The corresponding matrices for $x$ and $y$ bonds follow by cyclic permutation of the indices. Assuming   $|t_2'|\ll |t_1|,|t_2|,|t_3|$, we calculate the charge density response to first order in $t_2'$. In this approximation, we neglect the second-neighbor exchange interaction generated by perturbation theory at order $(t_2')^2$, keeping only  the nearest-neighbor exchange couplings as in Eq.~(\ref{eq:H_ext}).  

Following  Ref.~\cite{Pereira2020}, we write the effective charge imbalance operator as $\delta n_l = \sum_{(jk)} \delta n_{l,(jk)}$, 
where the sum over $(jk)$ runs  over   pairs of sites such that $jkl$ forms a triangle, and each triangle is counted  once.   These triangles contain two nearest-neighbor bonds and one next-nearest-neighbor bond, see the examples in Fig.~\ref{fig10}(b). The calculation of $\delta n_{l} $ requires the generator of the canonical transformation up to  second order in the hopping matrices, $S\approx S_1+S_2$.  After the projection  onto the $j_{\text{eff}}=\frac12$ subspace, we write the end result in the form 
\begin{equation}
\delta n_{l,(jk)}  = 
\sum_{\alpha\beta} \Big(  \mathcal{C}_{jkl}^{\alpha 0 \beta } \sigma_j^\alpha \sigma_l^\beta \!+\mathcal{C}_{jkl}^{ 0 \alpha\beta } \sigma_k^\alpha \sigma_l^\beta\! +  
\mathcal{C}_{jkl}^{   \alpha\beta 0 } \sigma_j^\alpha \sigma_k^\beta  \Big).\label{spincorr}
\end{equation}
Note that the effective density operator  involves only two-spin operators because it must be invariant under time reversal. The coefficients $ \mathcal{C}_{jkl}^{\mu\nu\rho }$ can be calculated as explained below. We find closed-form but lengthy expressions for general values of the hopping parameters. For $t_1=t_3=0$ and $t_2,t_2'\neq0$, we recover the result of  Ref.~\cite{Pereira2020}, in which the nonzero coefficients are diagonal in spin indices, e.g., $\mc C^{\alpha\beta0}_{jkl}\sim  \delta_{\alpha\beta} t_2^2t_2'/U^3$. Similarly to the derivation of the effective Hamiltonian, the addition of the subleading hopping parameters $t_1$ and $t_3$ generates off-diagonal terms in $\delta n_{l,(jk)}$ which are reminiscent of the $\Gamma$ interaction. 
Equation (\ref{spincorr}) implies that the charge density profile of a given state is determined by its spin correlations. Charge neutrality of the Mott insulator, $\sum_l\langle \delta n_l\rangle=0$, implies that there is no charge polarization in a homogeneous state where $\langle \delta n_l\rangle$ is uniform.  This condition is indeed satisfied when we impose that the spin correlations on different bonds respect translation and rotation symmetries, which provides a nontrivial check for the coefficients  $\mathcal{C}_{jkl}^{\mu\nu\rho }$. 

Let us outline some steps in the calculation of the coefficients $\mathcal{C}_{jkl}^{\mu\nu\rho }$ in Eq.~(\ref{spincorr}). 
At third order in the hopping term, the canonical transformation in Eq.~(\ref{transformdeltan}) gives
$\delta n_l^{(3)} =  -   S_2^{-}\,[  \bar N_l, S_1^{+} ]     + \text{h.c.}$, where
we organize the contributions in terms of triangles with site $l$ at one vertex.  In this notation, the contribution from each triangle with two other sites $(jk)\equiv(kj)$ contains two terms, $\delta n^{(3)}_{l,(jk)} = 
    \delta n^{(3)}_{l,jk} + 
    \delta n^{(3)}_{l,kj}$. Explicit  expressions for the  matrix elements of $\delta n^{(3)}_{l,(jk)}$ can be found in Ref. \cite{Pereira2020}.
 The last step is to project these matrices onto the  low-energy  subspace spanned by the states in Eq.~(\ref{jeff12}). The coefficients in  Eq.~(\ref{spincorr}) are given by 
\begin{equation}
    \mathcal{C}_{jkl}^{\mu\nu\rho} = \frac{1}{8} \text{Tr} \left( \mc P_{\rm low}  \delta n_{l,(jk)}^{(3)} \mc P_{\rm low} \sigma^\mu_j \sigma^\nu_k \sigma^\rho_l  \right)
, \label{eq:II-5-C}
\end{equation}
where $\sigma^{0}=\mathbb{1}$.
Since the charge density operator is even under time reversal, terms that act nontrivially on an odd number of spins vanish identically,  
\be
     \mathcal{C}_{jkl}^{\alpha\beta\gamma} =     \mathcal{C}_{jkl}^{\alpha00} =     \mathcal{C}_{jkl}^{0\alpha0}= \mathcal{C}_{jkl}^{00\alpha} = 0,
\ee
with $\alpha, \beta, \gamma \in \{1,2,3\}$.  
The nonzero terms can be written as in Eq.~(\ref{spincorr}) and depend on the specific triangle.  The simplest  coefficients are the ones that are already present in the solvable Kitaev model \cite{Pereira2020}. For instance, for the top triangle   in Fig.~\ref{fig10}(b), we obtain
\be
\mc{C}_{jkl}^{110}=
\mc{C}_{jkl}^{220}=
\frac{t_2^2   t_2^\prime }{U^3}\frac{\eta ^2 (1-2 \eta)}{9 (1-\eta)^3 (1-3 \eta )^3} ,
\ee
where $\eta=J_H/U<1/3$. Note that this term is sensitive to the sign of the second-neighbor hopping $t_2'$. In Ref.~\cite{Pereira2020}, the charge imbalance was calculated assuming a positive value of $t_2'$, but in this work we use $t_2'<0$ as obtained in Ref.~\cite{Winter_2016} for $\alpha$-RuCl$_3$. As an example for a coefficient associated with off-diagonal terms in $\delta n_l$, which are generated by the hoppings $t_1$ and $t_3$, we have
\begin{widetext} 
\be
\mc{C}_{jkl}^{120}=-\frac{ t_2^\prime }{U^3} \frac{\eta  (t_{1}-t_{3}) \left[\left(276 \eta ^4-94 \eta ^2-6 \eta +22\right) t_{1}+\left(26 \eta ^4-20 \eta ^3-7 \eta ^2-4 \eta +5\right) t_{3}\right]}{54 (1-\eta )^3 (1+2 \eta )^2 (1-3 \eta )^3}.
\ee
\end{widetext}

\section*{Discussion}

We have studied how vortices in Kitaev spin liquids generate and respond to nonuniform electric fields. While Kitaev materials are   Mott insulators, charge fluctuations can be generated at low energies by inhomogeneous spin correlations that carry signatures of localized excitations. To describe this effect, we started from the three-orbital Hubbard-Kanamori model for Kitaev materials.  Using a canonical transformation, we obtain effective operators in the low-energy sector in terms of spin operators that act on the pseudospin-$1/2$ states.  The effective spin Hamiltonian is the extended Kitaev model in a magnetic field, in which the exact solvability is broken by the  Heisenberg and  $\Gamma$  interactions as well as by a Zeeman coupling to a magnetic field. The effective density operator is generated at the level of third-order perturbation theory in the hopping terms. Generalizing the results of Ref. \cite{Pereira2020}, we found that the effective density operator associated with the extended Kitaev model contains all two-spin operators allowed by symmetry, including off-diagonal terms that are absent in the pure Kitaev model.

We have developed and applied a Majorana mean-field approach which allows to consider inhomogeneous parameters. 
While this approach is exact for the pure Kitaev model, we have demonstrated that it 
captures qualitative features of the Kitaev spin liquid phase in the extended $JK\Gamma$ model, where additional
spin interactions are present.  This model is believed to describe the candidate material $\alpha$-RuCl$_3$.
The electric charge distribution follows by computing the spin correlations around vortices in the mean-field approach.
Importantly, vortices remain localized on sufficiently long time scales even in the presence of small perturbations around the Kitaev limit, as long as the system remains deep in the Kitaev spin liquid phase.
   We find that the charge profile decays 
with the distance from the vortex in an oscillatory fashion.  

Our results allow us to calculate the
intrinsic electric quadrupole tensor of a vortex which is far away from all other vortices.  
The anisotropy of the quadrupole tensor can here be controlled by the magnetic field, 
and depending on the parameter regime, the interaction between different vortices is either repulsive or attractive.  
The interaction is generally enhanced by the $\Gamma$ interaction.  

Finally,
in the presence of local STM tips near vortices,
we find that one can close the vortex gap by applying a local electric potential to the tips.  We thus predict that
one can create vortices in a Kitaev spin liquid by means of STM tips in a controlled way.
Given the recent advances in STM technology \cite{Wagner2019,Bian2021,Yin2021}, 
our work paves the way for the electrical detection and manipulation in Kitaev materials. 
In particular, the successful control of Ising anyons in such materials would constitute a key step toward
 implementing a platform for topological quantum computation. 

\bibliography{references}

\section*{Data availability} 
All data supporting the findings of this paper are shown in the paper. 
Raw data and code used for preparing the figures are accessible on Zenodo under
\textcolor{blue}{https://doi.org/10.5281/zenodo.10616443}.

\section*{Acknowledgments}  
We acknowledge funding by the Brazilian agency CNPq and by the Coordena\c{c}\~{a}o de Aperfei\c{c}oamento de Pessoal de N{\'i}vel Superior - Brasil (CAPES), by the Deutsche Forschungsgemeinschaft (DFG, German Research Foundation), Projektnummer 277101999 - TRR 183 (project B04),  and under Germany's Excellence Strategy - Cluster of Excellence Matter and Light for Quantum Computing (ML4Q) EXC 2004/1 - 390534769. This work was supported by a grant from the Simons Foundation (1023171, R.G.P.). Research at IIP-UFRN is supported by Brazilian ministries MEC and MCTI.\\

\section*{Author contributions}
R.E. and R.G.P. conceived and managed the project.  The mean-field calculations were developed and carried out by L.R.D.F. with help from T.B.,
and the results were interpreted and discussed by all authors.  The paper was written by R.E. and R.G.P., with 
input from all authors. 

\section*{Competing interests}
The authors declare no competing interest.

 \appendix
 
\begin{figure*}[t]
\begin{center}
\includegraphics[width=\textwidth]{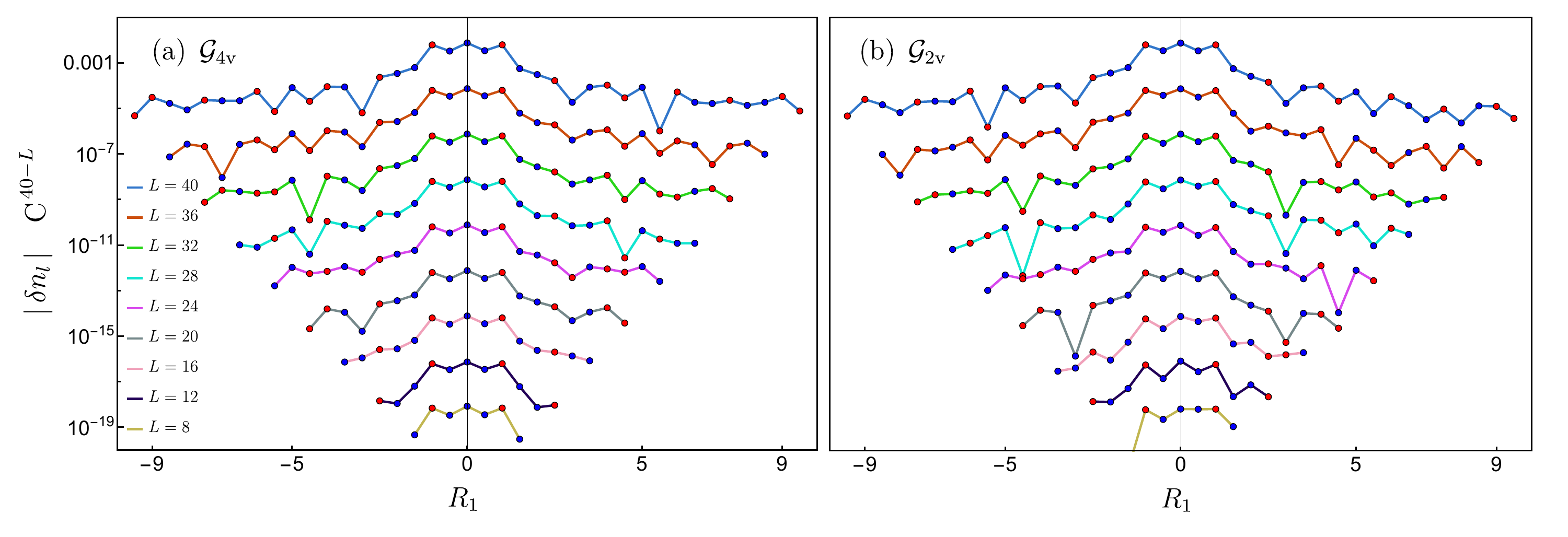}
\end{center}
\caption{ 
Charge distribution near a vortex for different system sizes for gauge configurations with (a) four, and (b) two maximally spaced vortices, along the path variable $R_1$ shown in Fig.~4 of the main text. Here we fix the parameters of the Hubbard-Kanamori model so that $\Gamma/|K|=0.3$, $J/|K|=-0.04$, and the magnetic field is $\mathbf{h}=0.2|K| \hat{\mathbf{z}}$. The curve for $L=40$ is shown using the same scale as in Fig.~5 of the main text. To aid visualization, the curves for other values of $L$ have been shifted down by rescaling the data by a factor $C^{40-L}$ with $C=1/\sqrt{10}$. 
}
\label{figS1}
\end{figure*}

\section{Finite-size effects}\label{sec1}

In Fig.~\ref{figS1} we show how finite-size effects impact the charge distribution around vortices in the extended Kitaev model. The coordinate $R_1$ corresponds to the zigzag path represented  in Fig.~4 of the main text.  In the thermodynamic limit and for infinitely separated vortices, the charge imbalance $\delta n_l$ must be symmetric with respect to the vortex center, i.e., with respect to $R_1\mapsto -R_1$. However, in a finite-size $L\times L$ geometry (periodic boundary conditions) with vortices located at maximal distance from each other, we see deviations from the symmetric distribution when the distance $|R_1|$ becomes comparable to the  separation between two vortices. The asymmetry in the charge distribution is substantially smaller for a configuration with four equally spaced vortices  than for two vortices. To see this, compare the data for smaller values of $L$ in Figs.~\ref{figS1}(a) and  Fig.~\ref{figS1}(b). This fact can be rationalized by noting that for periodic boundary conditions, the four-vortex configuration preserves a $C_3$ lattice rotation symmetry about the center of a given vortex, which helps to minimize finite-size effects. As we increase $L$, the results for both four-vortex and two-vortex configurations converge to the same values, especially close to the vortex, where the charge imbalance is larger. We have verified that all results reported in the main text
for the components of the quadrupole tensor, in particular the anisotropy parameter $\Delta Q$, are 
fully converged for $L\geq 40$.

\section{Vortex quadrupole moment for negative values of $\Gamma$}\label{sec1b}

\begin{figure}[t]
\begin{center}
\includegraphics[width=.95\columnwidth]{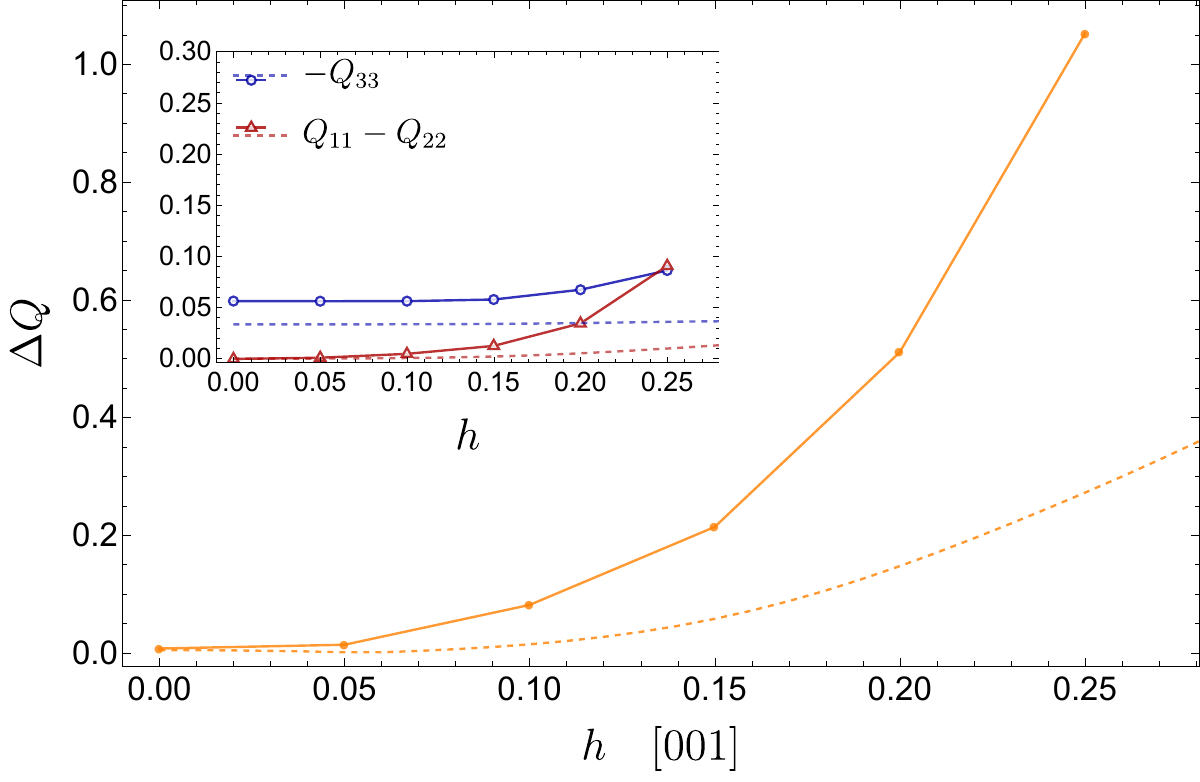}
\end{center}
\caption{
Same as Fig.~7 of the main text but for  $\Gamma = -0.3 |K|$, i.e., opposite sign of $\Gamma$. 
}
\label{figS2}
\end{figure}

When discussing the vortex charge density profile in the main text, we have focused on the parameter regime  $\Gamma>0$, which is relevant for $\alpha$-RuCl$_3$. In Fig.~\ref{figS2}, we show the anisotropy parameter $\Delta Q$ and other components of the quadrupole tensor for a negative value of $\Gamma$.  Comparing with Fig.~7 of the main text, we observe that the result is qualitatively similar to that for $\Gamma>0$.

\section{On the vortex lifetime}\label{sec2}

In our setup, vortices could be created by different mechanisms. Once they have been trapped by a local potential, they can decay if they meet another vortex that has been created far away in the bulk but then has propagated to the position of our designated vortex. 
As shown in Ref.~\cite{Joy_2022},  the time scale for two vortices to meet is given by 
$\tau_{VV}\sim 1/(Dn_V)$. Here $D$ is the diffusion constant, which is related to the vortex mobility $\mu$ by the Einstein relation, $D=\mu T$, and $n_V$ is the density of vortices (we set the Boltzmann constant $k_B=1$). At temperatures $T\ll \Delta_{\rm 2v}$,
where $\Delta_{\rm 2v}$ is the vison gap for creating a pair of vortices \cite{Savary2017,Joy_2022},
the diffusion constant becomes independent of the  vison dispersion and  is given by 
$D\approx 6v_m^2/T$, where $v_m$ is the characteristic velocity of Majorana fermion excitations. Since in this regime the diffusion constant does not depend explicitly on the effective hopping parameter of the visons, there is no strong dependence on the magnetic field, of arbitrary direction.
On the other hand, the vortex density becomes small at temperatures far below the vison gap, $T\ll \Delta_{\rm 2v}$.  For this reason, we expect the trapped vortex to have a very long lifetime at low temperatures. In the main text, we therefore assume that vortices are effectively stable and spatially localized entities.

\section{Discussion of Eq.~(14)}\label{sec3}

Unlike the other spin interactions included in our model, the DM interaction requires breaking bond inversion symmetries.  Starting from the Hubbard-Kanamori model, the DM interaction can be generated, for instance,  by generalizing the hopping matrix to be asymmetric, which introduces  three new hopping parameters; see e.g.  the derivation in Ref.~\cite{Winter2017rev}. In addition, the locally applied potential $V_0$ induces lattice distortions and introduces an anisotropic crystal field term of the form $\delta_{\rm CF} (\mb L_i\cdot \mb n)^2$, where  $\mb n$ is a unit vector in the direction of the local electric field and $\delta_{\rm CF}\sim V_0$ is the energy scale associated with the crystal field splitting. This term enters in the atomic Hamiltonian and modifies the expressions for the low-energy states in Eq.~(17) of the main text, which would now depend on the ratio between $\delta_{\rm CF}$ and the spin-orbit coupling $\lambda$. Instead of introducing a large number of new parameters in our model, we here prefer to follow a more phenomenological approach and include a single DM term which is allowed by symmetry, with a coupling constant that increases linearly with the local potential $V_0$. This dependence is plausible because the DM vector for nearest-neighbor bonds vanishes in the absence of the electric potential. We also fixed the dependence on the Hubbard interaction and Hund's coupling to be in terms of $\Delta E_1 = U-3J_H$ because this is the lowest among the three energy scales in Eq.~(20) of the main text, and it sets an upper bound for the electric potential that can be applied before the perturbative expressions break down. We note that this dependence
on $\Delta E_1$ does show up in some of the several interaction terms  generated by an asymmetric hopping matrix
(see Ref.~\cite{Winter2017rev}). 

In summary, we propose Eq.~(14) in the main text as the simplest expression that captures the main properties of the DM interaction and allows one to probe its effects on the vortex gap. A more detailed study, including all possible DM-type terms, should be guided by material-specific parameters as determined by {\it ab initio} calculations.

\end{document}